\documentclass[12pt]{article}
\usepackage{graphics}
\usepackage{amsmath, amssymb, graphics, setspace,booktabs}
\usepackage{comment}

\newcommand{\be}{\begin{equation}}
\newcommand{\ee}{\end{equation}}
\newcommand{\bea}{\begin{eqnarray}}
\newcommand{\eea}{\end{eqnarray}}
\newcommand{\ba}{\begin{array}}
\newcommand{\ea}{\end{array}}

\begin{document}
	%
	\vspace*{1.0cm}
	
	\begin{center}
		\baselineskip 20pt
		{\Large\bf
			Couplings in Renormalizable Supersymmetric SO(10) Models
		}
		\vspace{1cm}

		{\large
			Zhi-Yong Chen \footnote{ E-mail: chenzhiyongczy@pku.edu.cn},
			Da-Xin Zhang\footnote{ E-mail: dxzhang@pku.edu.cn}
			and
			Xian-Zheng Bai\footnote{E-mail: 15012100098@pku.edu.cn}
		}
		\vspace{.5cm}
		
		{\baselineskip 20pt \it
			School of Physics and State Key Laboratory of Nuclear Physics and Technology, \\
			Peking University, Beijing 100871, China}
		
		\vspace{.5cm}
		
		\vspace{1.5cm} {\bf Abstract}
	\end{center}
	We study the most general renormalizable couplings containing Higgs
	$H(10)$, $D(120)$, $\overline{\Delta}(\overline{126})+ \Delta(126)$, $A(45)$,
	$E(54)$ and $\Phi(210)$ in the supersymmetric SO(10) models.
	The Clebsch-Gordan Coefficients are calculated using the maximal subgroup $SU(5)\times U(1)_X$.
	
	\vspace{1em}
	
	\thispagestyle{empty}
	
	\newpage

\section{Introduction}

Supersymmetric (SUSY) Grand Unification Theories (GUTs)  of SO(10)\cite{so10a,so10b}
are very important candidates for the new physics
beyond the Standard Model (SM).
Since the SM gauge group $G_{321} = SU(3)_{C} \times SU(2)_{L} \times U(1)_{Y}$
is not a maximal subgroup of SO(10),
there are many different routines to break SO(10) into $G_{321}$ through
approximately intermediate symmetries.
Two maximal  subgroups $G_{422} = SU(4)_{C} \times SU(2)_{L} \times SU(2)_{R}$  and
$G_{51}=SU(5)\times U(1)_X$ are usually taken as the intermediate
symmetries. In practice, even without any intermediate symmetry breaking scale,
the  maximal  subgroups are used to distinguish different states which are in the same SO(10) representations
but also have the same SM representations.

Among the many systematic studies on the SUSY SO(10)
models\cite{he,lee,sato,np597,nath1,nath2,so10c,so10d,so10e,fuku,so10f,check1,check2,np711,np757,nath3,so10g,np857,np882,nath2015},
many of them  focus on the so-called Minimal SUSY SO(10) (MSSO10) \cite{so10c,so10d,so10e}
which contains the Higgs superfields  $H(10)$, $\overline{\Delta}(\overline{126})+ \Delta(126)$ and $\Phi(210)$.
The simplest extension of MSSO10 which contains one more $D(120)$ is studied in \cite{so10g,np857,np882}.
There are also models which use  $A(45)$ and $E(54)$ instead of $\Phi(210)$ to break SO(10)\cite{np597,lizhang}.
Most of the previous studies use the $G_{422}$ subgroup,
except in \cite{nath2015} where the MSSO10 is studied using the $G_{51}$.

Using $G_{422}$, the most general renormalizable couplings containing fields
$H(10)$, $D(120)$, $\overline{\Delta}(\overline{126})+ \Delta(126)$, $A(45)$,
$E(54)$ and $\Phi(210)$ have been studied in \cite{fuku}.
This general model offers powerful tools for  building realistic models
which may need all these fields \cite{cz2}.
However, the general model has not been checked entirely and an
independent  treatment using
$G_{51}$ subgroup is absent.
The realization of gauge coupling unification\cite{luo1,luo2,luo3,luo4}
can be most easily discussed within the SU(5) subgroup of SO(10),
even this SU(5) is only approximate.
Also, in the missing partner model of SO(10)\cite{mpmso10},
the solution to the outstanding problem of doublet-triplet splitting
depends on the group structure of SU(5).
The recent study of the $B-L=-2$ operators is also based on the SU(5) language\cite{nath2015}.
Thus a study of the most general renormalizable couplings containing fields
$H(10)$, $D(120)$, $\overline{\Delta}(\overline{126})+ \Delta(126)$, $A(45)$,
$E(54)$ and $\Phi(210)$
based on the $G_{51}$ subgroup is highly desirable,
both as an independent check on \cite{fuku} and as a  tool for future model building,
compliment to the $G_{422}$ approach.

In the present work, we will study the SUSY SO(10) models using its maximal  subgroup $G_{51}$.
Using the most general renormalizable superpotential as in \cite{fuku},
we will calculate the Clebsch-Gordan Coefficients (CGCs) using $G_{51}$.
After introducing notations and explaining the states in Section 2,
we will
give illustrative examples on the CGC calculations in Section 3,
study the  superpotential and the symmetry breaking conditions in Section 4,
and present all the masses and mass matrices in Section 5.
Then we will compare our results with those in the literature in Section 6 and summarize in Section 7.

\section{Notations and states}

A SO(2N) group can be studied using the SU(N) basis \cite{so2n}.
Our SO(10) notations mainly follow \cite{fuku} in the
SU(5) or the $Y$ diagonal basis\cite{sato} defined by
$1 \equiv 1' + 2'i$,
$2 \equiv 1' - 2'i$,
$\dots$,
$9 \equiv 9' + 0'i$,
$0 \equiv 9' - 0'i$,
besides a normalization factor $\frac{1}{\sqrt{2}}$.
Here $1', 2', \cdots, 9', 0'$ are the conventional  SO(10) basis.
Then
$(1, 3, 5, 7, 9)$ and
$(2, 4, 6, 8, 0)$ transform as $(\mathbf{5},2)$ and $(\mathbf{\overline{5}},-2)$,
respectively, under $SU(5) \times U(1)_X$.
Also, in larger SU(5) representations,
they  correspond to the superscripts and
 the subscripts, respectively, of tensors of higher ranks.
 We will take $1,2,3,4$ as the weak indices, and $6,7,8,9,0$ as the color indices.
 
For $H(10)$ of SO(10), it contains
\begin{eqnarray}\
	\widehat{H}_{(5, 2)}^{(1,2,\frac{1}{2})} &=& [1, 3], \qquad  \widehat{H}_{(5, 2)}^{(3,1, -\frac{1}{3})} = [5, 7, 9],\nonumber \\
\widehat{H}_{(\overline{5}, -2)}^{(1,2, -\frac{1}{2})} &=& [2, 4],	 \qquad \widehat{H}_{(\overline{5}, -2)}^{(3, 1, \frac{1}{3})} = [6, 8, 0].
\end{eqnarray}
Hereon we will denote a state as $R_{(I,x)}^i$, where $R$ and $I$ are the SO(10) and
SU(5) representations, respectively, $x$ is the $U(1)_X$ quantum number, and $i$ is the  SM representation.

The states in the anti-symmetric $A(45)$ of SO(10) can be constructed from the product of two (different) $H(10)$s.
$(5 \otimes \overline{5})_A = 1 \oplus 24$ gives
\begin{equation}
	\left(
	\begin{array}{c}
		1 \\
		3 \\
		5 \\
		7 \\
		9
	\end{array}
	\right) \otimes
	\left(
	\begin{array}{ccccc}
		2 & 4 & 6 & 8 & 0
	\end{array}
	\right)
-\left[ \left(
	\begin{array}{c}
		2 \\
		4 \\
		6 \\
		8 \\
		0
	\end{array}
	\right) \otimes
	\left(
	\begin{array}{ccccc}
		1 & 3 & 5 & 7 & 9
	\end{array}
	\right)\right]^T
	=
	\left(
	\begin{array}{cc|ccc}
		(12) & (14) & (16) & (18) & (10) \\
		(32) & (34) & (36) & (38) & (30) \\ \hline
		(52) & (54) & (56) & (58) & (50) \\
		(72) & (74) & (76) & (78) & (70) \\
		(92) & (94) & (96) & (98) & (90) \\
	\end{array}
	\right),
\end{equation}
so that
\begin{eqnarray}
	\widehat{A}^{(1,1,0)}_{(1, 0)} &=& \frac{i}{\sqrt{10}} (12 + 34 + 56 + 78 + 90 ), \nonumber \\
	 \widehat{A}^{(1,1,0)}_{(24, 0)} &=& \frac{i}{\sqrt{60}} \Big({\it 3}\times[12 + 34] - {\it 2}\times[56 + 78 + 90] \Big),  \nonumber \\
	 \widehat{A}^{(1,3,0)}_{(24, 0)} &=& \frac{i}{\sqrt{2}} \Big(14, 32,  \frac{12 - 34}{\sqrt{2}} \Big), \nonumber \\
	 \widehat{A}^{(8,1,0)}_{(24, 0)} &=& \frac{i}{\sqrt{2}}
	 \left(58,50,70,76,96,98,\frac{1}{\sqrt{2}}\left[56-78 \right],
	 \frac{1}{\sqrt{6}}\left[56+78-{\it 2}[90] \right] \right), \nonumber \\
	 \widehat{A}^{(3,2, -\frac{5}{6})}_{(24, 0)}
	 &=& \frac{i}{\sqrt{2}} (25, 27, 29, 45, 47, 49)
= \frac{i}{\sqrt{2}} \Big( \left[2,4 \right] \left[5,7,9 \right] \Big), \nonumber \\
	  \widehat{A}^{(\overline{3},2, \frac{5}{6})}_{(24, 0)} &=& -\frac{i}{\sqrt{2}} \Big( \left[1, 3 \right] \left[6,8,0 \right] \Big),
\end{eqnarray}
where the parenthesis means anti-symmetrization $(ab) = ab - ba$, and the square bracket are used for grouping of indices into complete
$SU(2)_L$ and $SU(3)_C$ representations.
In the brackets, the italicized integer numbers stand for
numerical factors to be distinguished from the SO(10) indices.
The anti-symmetric part of the SU(5) product
$(5 \otimes 5)_{A} = 10$ gives
\begin{equation}
	\left(
	\begin{array}{c}
	1 \\
	3 \\
	5 \\
	7 \\
	9
	\end{array}
	\right) \otimes
	\left(
	\begin{array}{ccccc}
	1 & 3 & 5 & 7 & 9
	\end{array}
	\right)
-\left[\left(
	\begin{array}{c}
	1 \\
	3 \\
	5 \\
	7 \\
	9
	\end{array}
	\right) \otimes
	\left(
	\begin{array}{ccccc}
	1 & 3 & 5 & 7 & 9
	\end{array}
	\right)\right]^T
	=
	\left(
		\begin{array}{cc|ccc}
		0 & (13) & (15) & (17) & (19) \\
		(31) & 0 & (35) & (37) & (39) \\ \hline
		(51) & (53) & 0 & (57) & (59) \\
		(71) & (73) & (75) & 0 & (79) \\
		(91) & (93) & (95) & (97) & 0 \\
		\end{array}
	\right),
\end{equation}
so that
\begin{eqnarray}
	\widehat{A}^{(1,1,1)}_{(10, 4)}
	&=& \frac{i}{\sqrt{2}} \left(13 \right),  \nonumber \\
	\widehat{A}^{(3,2,\frac{1}{6})}_{(10, 4)}
	&=& \frac{i}{\sqrt{2}} (15, 17, 19, 35, 37, 39)
= \frac{i}{\sqrt{2}} \Big( [1,3][5, 7, 9] \Big), \nonumber \\
	\widehat{A}^{(\overline{3},1, -\frac{2}{3})}_{(10, 4)}
	&=& \frac{i}{\sqrt{2}} (79, 95, 57),
\end{eqnarray}
and
$(\overline{5} \otimes \overline{5})_{A} = \overline{10}$ gives
\begin{equation}
	\left(
	\begin{array}{c}
	2 \\
	4 \\
	6 \\
	8 \\
	0
	\end{array}
	\right) \otimes
	\left(
	\begin{array}{ccccc}
	2 & 4 & 6 & 8 & 0
	\end{array}
	\right)
-\left[	\left(
	\begin{array}{c}
	2 \\
	4 \\
	6 \\
	8 \\
	0
	\end{array}
	\right) \otimes
	\left(
	\begin{array}{ccccc}
	2 & 4 & 6 & 8 & 0
	\end{array}
	\right)\right]^T
	=
	\left(
		\begin{array}{cc|ccc}
		0 & (24) & (26) & (28) & (20) \\
		(42) & 0 & (46) & (48) & (40) \\ \hline
		(62) & (64) & 0 & (68) & (60) \\
		(82) & (84) & (86) & 0 & (80) \\
		(02) & (04) & (06) & (08) & 0 \\
		\end{array}
	\right),
\end{equation}
so that
\begin{eqnarray}
	\widehat{A}^{(1,1,1)}_{(\overline{10}, -4)}
	&=& -\frac{i}{\sqrt{2}} \left(24 \right),  \nonumber \\
	\widehat{A}^{(\overline{3},2, -\frac{1}{6})}_{(\overline{10}, -4)}
	&=& -\frac{i}{\sqrt{2}} (26, 28, 20, 46, 48, 40)
= -\frac{i}{\sqrt{2}} \Big( [2,4][6, 8, 0] \Big) \nonumber, \\
	\widehat{A}^{(3,1, \frac{2}{3})}_{(\overline{10}, -4)}
	&=& -\frac{i}{\sqrt{2}} (80, 06, 68).
\end{eqnarray}
In (3,5,7) the factor $i$ or $-i$ are included in accord with the notations in \cite{fuku}.
The states in $E(54)$ of SO(10) can be constructed in a similar way,
{\it e.g.},
\begin{eqnarray}
	\widehat{E}^{(1,1,0)}_{(24, 0)} &=& \frac{1}{\sqrt{60}} \{ {\it 3}\times[12 + 34] - {\it 2}\times[56 + 78 + 90] \}.  \end{eqnarray}
Here the curly brackets stand for symmetrization $\{ab\} = ab + ba$.

Higher ranked representations of SO(10) can be constructed similarly.
$D(120)$ can be constructed from anti-symmetrizing $10 \otimes 45$,
and $\Delta (126)-\overline{\Delta}(\overline{126})$ from $45 \otimes 120$.
For example,
\begin{eqnarray}
	\widehat{D}^{(1,1, -1)}_{(\overline{10}, 6)}
	&=& \frac{1}{N} \Big(\widehat{H}_{(5, 2)}^{(3, 1, -\frac{1}{3})} \otimes \widehat{A}_{(10, 4)}^{(\overline{3}, 1, -\frac{2}{3})}\Big)_A
	=  \frac{1}{N}\Big([5, 7, 9] \otimes (79, 95, 57) \Big)_A
	= \frac{1}{\sqrt{6}}(579), \nonumber \\
	 \widehat{\overline{\Delta}}^{(1,1,0)}_{(1,10)}
	  &=&
	  \frac{1}{N^\prime}\Big(\widehat{D}^{(1,1, -1)}_{(\overline{10}, 6)} \otimes \widehat{A}^{(1,1,1)}_{(10, 4)} \Big)_A
		=	\frac{1}{N^\prime} \Big( (579) \otimes (13) \Big)_A
		=  \frac{1}{\sqrt{120}} (13579),
\end{eqnarray}
where $N, N^\prime$ are normalization factors which are only determined at the ends of the identities.
Under the complex conjugation (c.c), we have
$\overline{1} = 2, \,
\overline{3} = 4, \,
\overline{5} = 6, \,
\overline{7} = 8, \,
\overline{9} = 0, $ and vice versa.
So, we have
\begin{eqnarray}
	\widehat{\Delta}^{(1,1,0)}_{(1, - 10)} = \frac{1}{\sqrt{120}} (\overline{13579}) = \frac{1}{\sqrt{120}}  (24680).
\end{eqnarray}
As can be easily verified, $\widehat{\overline{\Delta}}^{(1,1,0)}_{(1,10)}$ and
$\widehat{\Delta}^{(1,1,0)}_{(1, - 10)}$ satisfy
\begin{eqnarray}
	i \varepsilon_{\bar{a}_1\bar{a}_2\bar{a}_3\bar{a}_4\bar{a}_5
		\bar{a}_6\bar{a}_7\bar{a}_8\bar{a}_9\bar{a}_{0}}
	\overline{\Delta}_{a_6a_7a_8a_9a_{0}} &=&
	\overline{\Delta}_{\bar{a}_1\bar{a}_2\bar{a}_3\bar{a}_4\bar{a}_5},
	\nonumber \\
	i \varepsilon_{\bar{a}_1\bar{a}_2\bar{a}_3\bar{a}_4\bar{a}_5
		\bar{a}_6\bar{a}_7\bar{a}_8\bar{a}_9\bar{a}_{0}}
	\Delta_{a_6a_7a_8a_9a_{0}} &=&
	-\Delta_{\bar{a}_1\bar{a}_2\bar{a}_3\bar{a}_4\bar{a}_5},
\end{eqnarray}
where $i \varepsilon_{1234567890} = 1$.

$\Phi(210)$ can be constructed from $45 \otimes 45$, {\it e.g.},
\begin{eqnarray}
\hat{\Phi}^{(1,1,0)}_{(1,0)} &=& \frac{1}{N_1}\widehat{A}^{(1,1,0)}_{(1,0)}\otimes \widehat{A}^{(1,1,0)}_{(1,0)} \nonumber \\
&=& -\frac{1}{\sqrt{240}}(1234 + 1256 + 1278 + 1296 +3456 + 3478 +3496 + 5678 + 5690 + 7890 ) \nonumber \\
&=& -\frac{1}{\sqrt{240}}(1234 + [5678 + 5690 + 7890] + [12 + 34][56 + 78 + 90]),  \\
\hat{\Phi}^{(1,1,0)}_{(24,0)} &=& \frac{1}{N_2}\widehat{A}^{(1,1,0)}_{(1, 0)} \otimes \widehat{A}^{(1,1,0)}_{(24, 0)}\nonumber \\
&=& - \frac{1}{N_2}([12 + 34] + [56 + 78 + 90]) \otimes ({\it 3}\times[12 + 34] - {\it 2}\times[56 + 78 + 90]) \nonumber \\
&=& - \frac{1}{\sqrt{90\times 24}}({\it 6}\times [1234] + [12 + 34][56+ 78 + 90] - {\it 4} \times [5678 + 5690 + 7890]), \nonumber \\
\end{eqnarray}
where $N_{1,2}$ are normalization factors,
and
\begin{equation}
	\widehat{A}^{(1,1,0)}_{(24,0)}\otimes \widehat{A}^{(1,1,0)}_{(24,0)}
	= ({\it 9}\times [1234] -{\it 6} [12 + 34][56+ 78 + 90] + {\it 4} \times [5678 + 5690 + 7890]) \nonumber
\end{equation}
contains not only $\hat{\Phi}^{(1,1,0)}_{(1,0)}$ and $\hat{\Phi}^{(1,1,0)}_{(24,0)}$,
but also a state orthogonal to them which is
\begin{eqnarray}
	\hat{\Phi}^{(1,1,0)}_{(75,0)} &=&
	- \frac{1}{\sqrt{18\times 24}}({\it3}\times [1234] - [12 + 34][56+ 78 + 90] +  [5678 + 5690 + 7890]). \nonumber \\
\end{eqnarray}

The SM singlets which break SO(10) into its SM subgroup when they get VEVs are:
    \begin{eqnarray}
		\widehat{a_1} &\equiv& \widehat{A}^{(1,1,0)}_{(1,0)} = \frac{i}{\sqrt{10}} (12 + 34 + 56 + 78 + 90),  \label{vev}\\
		\widehat{a_2} &\equiv& \widehat{A}^{(1,1,0)}_{(24, 0)} = \frac{i}{\sqrt{60}} ({\it 3}\times[12 + 34] - {\it 2}\times[56 + 78 + 90]),  \\
		\widehat{E} &\equiv& \widehat{E}^{(1,1,0)}_{(24,0)}= \frac{1}{\sqrt{60}} \{{\it 3}\times[12 + 34] - {\it 2}\times[56 + 78 + 90]\}, \\
		\widehat{V_R} &\equiv& \widehat{\Delta}^{(1,1,0)}_{(1,- 10)}  = \frac{1}{\sqrt{120}}(24680),\\
		\widehat{\overline{V_R}} &\equiv& \widehat{\overline{\Delta}}^{(1,1,0)}_{(1,10)}  = \frac{1}{\sqrt{120}}(13579),\\
		\widehat{\phi}_1 &\equiv& \widehat{\Phi}^{(1,1,0)}_{(1,0)} = -\frac{1}{\sqrt{240}}(1234 + [5678 + 5690 + 7890] + [12 + 34][56 + 78 + 90]),  \\
		\widehat{\phi}_2 &\equiv& \widehat{\Phi}^{(1,1,0)}_{(24,0)} =- \frac{1}{\sqrt{90\times 24}}({\it 6}\times [1234] - {\it 4} \times [5678 + 5690 + 7890] + [12 + 34][56+ 78 + 90] ), \nonumber \\
		\\
		\widehat{\phi}_3 &\equiv& \widehat{\Phi}^{(1,1,0)}_{(75,0)} = - \frac{1}{\sqrt{18 \times 24}}({\it 3} \times [1234]
		+ [5678 + 5690 + 7890]
		- [12 + 34][56 + 78 + 90] ). \nonumber \\
	\end{eqnarray}
A minus sign is included in (12-14) following \cite{fuku}.
The normalizaion factors before are chosen according to the requirement of
$\widehat{R}_{(I,x)}^i \widehat{R}_{(\overline{I}, -x)}^{\overline{i}} = 1$.
Note that the extra factors $\pm i$ in $A$ and $-1$ in $\phi_{1,2,3}$  follow \cite{fuku},
so that there may exist extra minus sign in some CGCs involving them which actually do not
violate the symmetries in the CGCs.

The other states are summarized in Table \ref{tabw}-\ref{tabo2},
where different states of the same SM representations in a same SO(10) field
are distinguished by their different representations under $G_{51}$.
In Table \ref{tabw}, all would-be Goldstone states are given.

\begin{table}[p]
\caption{States in the would-be Goldstone modes.}
\label{tabw}
\begin{center}
\begin{tabular}{|c|c|c|}
\hline \hline
${\bf(1,1},{1}) +c.c.$
& $\widehat{A}^{(1,1,1)}_{(10, 4)}$
& $\frac{i}{\sqrt{2}} \left(13 \right) +c.c. $
\\
& $\widehat{D}^{(1,1,1)}_{(10, -6)}$
& $\frac{1}{\sqrt{6}} \left(680 \right) +c.c. $
\\
& $\widehat{\Delta}^{(1,1,1)}_{(10, -6)}$
& $\frac{1}{\sqrt{240}} \left( \left[12+34 \right]  680 \right) +c.c. $
\\
& $\widehat{\Phi}^{(1,1,1)}_{(10, 4)}$
& $\frac{1}{\sqrt{72}} (13  \left[56+78+90 \right]) +c.c. $
\\
\hline
${\bf(3,1},{\frac{2}{3}}) +c.c.$
& $\widehat{A}^{(3,1,\frac{2}{3})}_{(\overline{10}, -4)}$
& $\frac{i}{\sqrt{2}} \left(80,06,68 \right) +c.c.$
\\
& $\widehat{D}^{(3,1,\frac{2}{3})}_{(\overline{10}, 6)}$
& $\frac{1}{\sqrt{6}} (13 \left[5,7,9 \right])+c.c.$
\\
& $\widehat{\overline{\Delta}}^{(3,1,\frac{2}{3})}_{(\overline{10}, 6)}$
& $\frac{1}{\sqrt{240}} \left( 13 \left[5 \left[78+90 \right],
7 \left[56+90 \right], 9 \left[56+78 \right] \right]\right)+c.c.$
\\
& $\widehat{\Phi}^{(3,1,\frac{2}{3})}_{(\overline{10}, -4)}$
& $\frac{1}{\sqrt{72}} (\left[12 + 34 \right] \left[80,06,68 \right] + \left[5680,7806,9068 \right]) + c.c. $
\\
& $\widehat{\Phi}^{(3,1,\frac{2}{3})}_{(40, -4)}$
& $\frac{1}{\sqrt{144}} ( {\it 2} \times \left[5680,7806,9068 \right] - \left[12 + 34 \right] \left[80,06,68 \right]) + c.c.$
\\
\hline
${\bf(3,2},{-\frac{5}{6}}) +c.c.$
& $\widehat{A}^{(3,2,-\frac{5}{6})}_{(24, 0)}$
& $\frac{i}{\sqrt{2}} \left(\left[2,4 \right] \left[5,7,9 \right] \right)+c.c.$
\\
& $\widehat{E}^{(3,2,-\frac{5}{6})}_{(24, 0)}$
& $\frac{1}{\sqrt{2}} \left\{\left[2,4 \right] \left[5,7,9\right] \right\}+c.c.$
\\
& $\widehat{\Phi}^{(3,2,-\frac{5}{6})}_{(24, 0)}$
& $\frac{1}{\sqrt{72}} \Big( \left[234,124 \right] \left[5,7,9 \right] $
\\
&& $+  \left[2,4 \right]
\left[5 \left[78+90 \right],
7 \left[56+90 \right],9 \left[56+78 \right] \right]\Big)+c.c. $
\\
& $\widehat{\Phi}^{(3,2,-\frac{5}{6})}_{(75, 0)}$
& $\frac{1}{\sqrt{144}} \Big( {\it 2} \times \left[234,124 \right] \left[5,7,9 \right]$ \\
&&
$- \left[2,4 \right]
\left[5 \left[78+90 \right],
7 \left[56+90 \right],9 \left[56+78 \right] \right]\Big)+c.c.$
\\
\hline
${\bf(3,2},{\frac{1}{6}}) +c.c.$
& $\widehat{A}^{(3,2,\frac{1}{6})}_{(10, 4)}$
& $\frac{i}{\sqrt{2}} (\left[1,3 \right] \left[5,7,9 \right])+c.c.$
\\
& $\widehat{E}^{(3,2,\frac{1}{6})}_{(15, 4)}$
& $\frac{1}{\sqrt{2}} \left\{\left[1,3 \right] \left[5,7,9 \right] \right\}+c.c.$
\\
& $\widehat{D}^{(3,2,\frac{1}{6})}_{(10, -6)}$
& $\frac{1}{\sqrt{6}} \left( \left[-4,2 \right] \left[80,06,68 \right] \right)+c.c.$
\\
& $\widehat{\Delta}^{(3,2,\frac{1}{6})}_{(10, -6)}$
& $\frac{1}{\sqrt{240}} \Big[
\left( \left[-124,234 \right]
\left[80,06,68 \right] \right)
+  \left(\left[-4,2 \right]
\left[5680,7806,9068 \right] \right)
\Big] +c.c. $
\\
& $\widehat{\overline{\Delta}}^{(3,2,\frac{1}{6})}_{(15, -6)}$
& $\frac{1}{\sqrt{240}} \Big[
\left( \left[-124,234 \right]
\left[80,06,68 \right] \right)
-  \left(\left[-4,2 \right]
\left[5680,7806,9068 \right] \right)
\Big] +c.c. $
\\
& $\widehat{\Phi}^{(3,2,\frac{1}{6})}_{(10, 4)}$
& $\frac{1}{\sqrt{72}} \Big( \left[134,123 \right] \left[5,7,9 \right]$ \\
&& $+  \left[1,3 \right]
\left[5 \left[78+90 \right],
7 \left[56+90 \right], 9 \left[56+78 \right] \right] \Big)+c.c. $
\\
& $\widehat{\Phi}^{(3,2,\frac{1}{6})}_{(\overline{40}, 4)}$
& $\frac{1}{\sqrt{144}}\Big( {\it 2}\times \left[134,123 \right] \left[5,7,9 \right]$
\\
&& $-  \left[1,3 \right]
\left[5 \left[78+90 \right],
7 \left[56+90 \right], 9 \left[56+78 \right] \right] \Big)+c.c.$
\\
\hline \hline
\end{tabular}
\end{center}
\end{table}
\begin{table}[p]
\caption{States in [${\bf(1,2},{\frac{1}{2}}) +c.c.$].}
\label{tabd}
\begin{center}
\begin{tabular}{|c|c|c|}
\hline \hline
${\bf(1,2},{\frac{1}{2}}) +c.c.$
& $\widehat{H}^{(1,2,\frac{1}{2})}_{(5, 2)}$
& $\left[1,3 \right] + c.c. $
\\
& $\widehat{D}^{(1,2,\frac{1}{2})}_{(5, 2)}$
& $\frac{1}{\sqrt{24}} \left(\left[134,123\right]  +  \left[1,3 \right]\left[56+78+90 \right]\right)+c.c.$
\\
& $\widehat{D}^{(1,2,\frac{1}{2})}_{(45, 2)}$
& $\frac{1}{\sqrt{72}} \left({\it 3 \times}\left[134,123\right]  -  \left[1,3 \right]\left[56+78+90 \right]\right)+c.c.$
\\
& $\widehat{\Delta}^{(1,2,\frac{1}{2})}_{(45, 2)}$
& $\frac{1}{\sqrt{720}} \Big[
   \left( \left[134,123 \right] \left[56+78+90 \right] \right)
-  \left( \left[1,3 \right] \left[5678+5690+7890 \right] \right) \Big] +c.c. $
\\
& $\widehat{\overline{\Delta}}^{(1,2,\frac{1}{2})}_{(5, 2)}$
& $\frac{1}{\sqrt{720}} \Big[
   \left( \left[134,123 \right] \left[56+78+90 \right] \right)
+  \left( \left[1,3 \right] \left[5678+5690+7890 \right] \right) \Big] +c.c. $
\\
& $\widehat{\Phi}^{(1,2,\frac{1}{2})}_{(5, -8) }$
& $\frac{1}{\sqrt{24}} \left( \left[-4,2 \right] 680 \right)+c.c. $
\\
\hline \hline
\end{tabular}
\end{center}
\end{table}
\begin{table}[p]
\caption{States in [${\bf(3,1},{-\frac{1}{3}}) +c.c.$].}
\label{tabt}
\begin{center}
\begin{tabular}{|c|c|c|}
\hline \hline
${\bf(3,1},{-\frac{1}{3}}) +c.c.$
& $\widehat{H}^{(3,1,-\frac{1}{3})}_{(5, 2)}$
& $\left[5,7,9 \right]+c.c.$
\\
& $\widehat{D}^{(3,1,-\frac{1}{3})}_{(5, 2)}$
& $\frac{1}{\sqrt{24}}\left( \left[12+34 \right] \left[5,7,9 \right]
+ \left[5\left[78+90 \right], 7\left[56+90 \right], 9\left[56+78 \right] \right] \right)+c.c.$
\\
& $\widehat{D}^{(3,1,-\frac{1}{3})}_{(\overline{45}, -2)}$
& $\frac{1}{\sqrt{24}}
\left( \left[12+34 \right] \left[5,7,9 \right]
 - \left[5\left[78+90 \right], 7\left[56+90 \right], 9\left[56+78 \right] \right] \right)+c.c.$
\\
& $\widehat{\Delta}^{(3,1,-\frac{1}{3})}_{(45, 2)}$
& $\frac{1}{\sqrt{240}} \Big[ \left(1234 \left[5,7,9 \right] \right)
-  \left(57890,56790,56789 \right) \Big] +c.c. $
\\
& $\widehat{\overline{\Delta}}^{(3,1,-\frac{1}{3})}_{(5, 2)}$
& $\frac{1}{\sqrt{720}} \Big[ \left(1234 \left[5,7,9 \right] \right)
+  \left(57890,56790,56789 \right)$
\\
&&
$+  \left( \left[12+34 \right]
\left[5\left[78+90 \right], 7\left[56+90 \right], 9\left[56+78 \right] \right] \right) \Big] +c.c. $
\\
& $\widehat{\overline{\Delta}}^{(3,1,-\frac{1}{3})}_{(50, 2)}$
&$\frac{1}{\sqrt{1440}} \Big[ {\it 2} \times \left(\left(1234 \left[5,7,9 \right] \right)
+  \left(57890,56790,56789 \right) \right) $
\\
&&
 $  - \left( \left[12+34 \right]
\left[5\left[78+90 \right], 7\left[56+90 \right], 9\left[56+78 \right] \right] \right)
 \Big] +c.c. $
\\
& $\widehat{\Phi}^{(3,1,-\frac{1}{3})}_{(5, -8)}$
&$\frac{1}{\sqrt{24}} \left(24 \left[80,06,68 \right] \right)+c.c. $
\\
\hline \hline
\end{tabular}
\end{center}
\end{table}
\begin{table}[p]
\caption{States in the other $G_{321}$ multiplets including $\Delta$
or $\overline{\Delta}$.}
\label{tabo1}
\begin{center}
\begin{tabular}{|c|c|c|}
\hline \hline
${\bf(1,1},{2}) +c.c.$
& $\widehat{\Delta}^{(1,1,2)}_{(\overline{50}, -2)}$
& $\frac{1}{\sqrt{120}} \left(13680 \right)+c.c. $
\\ \hline
${\bf(1,3},{1}) +c.c.$
& $\widehat{E}^{(1,3,1)}_{(15, 4)}$
& $\left[11,-33,-\frac{\left\{13 \right\}}{\sqrt{2}} \right]+c.c$
\\
& $\widehat{\overline{\Delta}}^{(1,3,1)}_{(15, -6)}$
& $\frac{1}{\sqrt{120}} \left(\left[14,32, \frac{12-34}{\sqrt{2}} \right]
 680 \right) +c.c.$
\\ \hline
${\bf(3,1},{-\frac{4}{3}}) +c.c.$
& $\widehat{D}^{(3,1,-\frac{4}{3})}_{(\overline{45}, -2)}$
& $\frac{1}{\sqrt{6}} \left( 24 \left[5,7,9 \right] \right)+c.c.$
\\
& $\widehat{\overline{\Delta}}^{(3,1,-\frac{4}{3})}_{(\overline{45}, -2)}$
& $\frac{1}{\sqrt{240}} \left( 24
\left[5 \left[78+90 \right],7 \left[56+90 \right],9 \left[56+78 \right] \right]
\right)
+c.c.$
\\ \hline
${\bf(3,2},{\frac{7}{6}}) +c.c.$
& $\widehat{D}^{(3,2,\frac{7}{6})}_{(\overline{45}, -2)}$
& $\frac{1}{\sqrt{6}} \left( \left[1,3 \right]
 \left[80, 06, 68 \right] \right) +c.c.$
\\
& $\widehat{\Delta}^{(3,2,\frac{7}{6})}_{(\overline{50}, -2)}$
& $\frac{1}{\sqrt{240}} \left( \left[134,123 \right] \left[80,06,68 \right]
- \left[1,3 \right] \left[5680, 7806, 9068 \right] \right)$ +c.c
\\
& $\widehat{\overline{\Delta}}^{(3,2,\frac{7}{6})}_{(\overline{45}, -2)}$
& $\frac{1}{\sqrt{240}} \left( \left[134,123 \right] \left[80,06,68 \right]
+ \left[1,3 \right]\left[5680, 7806, 9068 \right] \right)$ +c.c
\\ \hline
${\bf(3,3},{-\frac{1}{3}}) +c.c.$
& $\widehat{D}^{(3,3,-\frac{1}{3})}_{(45, 2)}$
& $\frac{1}{\sqrt{6}} \left( \left[14,32,\frac{12-34}{\sqrt{2}} \right]
\left[5,7,9 \right] \right) +c.c.$
\\
& $\widehat{\Delta}^{(3,3,-\frac{1}{3})}_{(45, 2)}$
& $\frac{1}{\sqrt{240}} \left( \left[14,32, \frac{12-34}{\sqrt{2}} \right] \right.$
\\
&& $\left. \left[5 \left[78+90 \right],7 \left[56+90 \right],9 \left[56+78 \right]
 \right] \right)
+c.c.$
\\ \hline
${\bf(6,1},{-\frac{2}{3}}) +c.c.$
& $\widehat{E}^{(6,1,-\frac{2}{3})}_{(15, 4)}$
& $\left[55,77,99,
      \frac{\{79\}}{\sqrt{2}},\frac{\{95\}}{\sqrt{2}},\frac{\{57\}}{\sqrt{2}} \right] +c.c.$
\\
& $\widehat{\overline{\Delta}}^{(6,1,-\frac{2}{3})}_{(15, -6)}$
& $\frac{1}{\sqrt{120}} \left( 24
\left[580, 670,689,
\frac{6 \left[90-78 \right]}{\sqrt{2}},
\frac{8 \left[56-90 \right]}{\sqrt{2}},
\frac{0 \left[78-56 \right]}{\sqrt{2}} \right] \right) +c.c.$
\\ \hline
${\bf(6,1},{\frac{1}{3}}) +c.c.$
& $\widehat{D}^{(6,1,\frac{1}{3})}_{(\overline{45}, -2)}$
& $\frac{1}{\sqrt{6}} \left(580, 670, 689,
\frac{6 \left[90-78 \right]}{\sqrt{2}},
\frac{8 \left[56-90 \right]}{\sqrt{2}},
\frac{0 \left[78-56 \right]}{\sqrt{2}} \right)+c.c.$
\\
& $\widehat{\overline{\Delta}}^{(6,1,\frac{1}{3})}_{(\overline{45}, -2)}$
& $\frac{1}{\sqrt{240}} \left( \left[12+34 \right]
\left[580, 670, 689,
\frac{6 \left[90-78 \right]}{\sqrt{2}},
\frac{8 \left[56-90 \right]}{\sqrt{2}},
\frac{0 \left[78-56 \right]}{\sqrt{2}} \right] \right)+c.c.$
\\ \hline
${\bf(6,1},{\frac{4}{3}}) +c.c.$
& $\widehat{\overline{\Delta}}^{(6,1,\frac{4}{3})}_{(50, 2)}$
& $\frac{1}{\sqrt{120}} \left(13
\left[580, 670, 689,
\frac{6 \left[90-78 \right]}{\sqrt{2}},
\frac{8 \left[56-90 \right]}{\sqrt{2}},
\frac{0 \left[78-56 \right]}{\sqrt{2}} \right] \right)+c.c.$
\\ \hline
${\bf(6,3},{\frac{1}{3}}) +c.c.$
& $\widehat{\Delta}^{(6,3,\frac{1}{3})}_{(\overline{50}, -2)}$
& $\frac{1}{\sqrt{120}} \left( \left[14,32, \frac{12-34}{\sqrt{2}} \right] \right.$
\\
&& $\left. \left[580, 670, 689,
\frac{6 \left[90-78 \right]}{\sqrt{2}},
\frac{8 \left[56-90 \right]}{\sqrt{2}},
\frac{0 \left[78-56 \right]}{\sqrt{2}} \right] \right)+c.c.$
\\ \hline
${\bf(8,2},{\frac{1}{2}}) +c.c.$
& $\widehat{D}^{(8,2,\frac{1}{2})}_{(45, 2)}$
& $\frac{1}{\sqrt{6}} \left( \left[1,3 \right] \left[58, 50, 70, 76, 96, 98,
\frac{56-78}{\sqrt{2}}, \frac{56+78-{\it 2}[90]}{\sqrt{6}} \right] \right)+c.c.$
\\
& $\widehat{\Delta}^{(8,2,\frac{1}{2})}_{(45, 2)}$
& $\frac{1}{\sqrt{240}} \left(
\left[134,123 \right] \left[58, 50, 70, 76, 96, 98,
\frac{56-78}{\sqrt{2}}, \frac{56+78-{\it 2}[90]}{\sqrt{6}} \right] \right.$
\\
&& $ +    \left[1,3 \right]
\left[5890, 5078, 7056, 7690, 9678, 9856, \right. $
\\
&& $\left.\left. \frac{5690-7890}{\sqrt{2}},
\frac{{\it 2}[5678]-5690-7890}{\sqrt{6}} \right] \right) + c.c.$
\\
& $\widehat{\overline{\Delta}}^{(8,2,\frac{1}{2})}_{(50, 2)}$
& $\frac{1}{\sqrt{240}} \left(
\left[134,123 \right] \left[58, 50, 70, 76, 96, 98,
\frac{56-78}{\sqrt{2}}, \frac{56+78-{\it 2}[90]}{\sqrt{6}} \right] \right.$
\\
&& $ -    \left[1,3 \right]
\left[5890, 5078, 7056, 7690, 9678, 9856, \right. $
\\
&& $\left.\left. \frac{5690-7890}{\sqrt{2}},
\frac{{\it 2}[5678]-5690-7890}{\sqrt{6}} \right] \right) + c.c.$
\\
\hline \hline
\end{tabular}
\end{center}
\end{table}
\begin{table}[p]
\caption{States in the other $G_{321}$ multiplets including $\Phi$.}
\label{tabo2}
\begin{center}
\begin{tabular}{|c|c|c|}
\hline \hline
${\bf(1,2},{\frac{3}{2}}) +c.c.$
& $\widehat{\Phi}^{(1,2,\frac{3}{2})}_{(40, -4)}$
& $\frac{1}{\sqrt{24}} \left( \left[1,3 \right]  680 \right)+c.c.$
\\
\hline
${\bf(1,3},{0})$
& $\widehat{A}^{(1,3,0)}_{(24, 0)}$
& $\frac{i}{\sqrt{2}} \left(14,32, \frac{12-34}{\sqrt{2}} \right)$
\\
& $\widehat{E}^{(1,3,0)}_{(24, 0)}$
& $\frac{1}{\sqrt{2}} \left\{14,32, \frac{12-34}{\sqrt{2}} \right\}$
\\
&$\widehat{\Phi}^{(1,3,0)}_{(24, 0)}$
&$\frac{1}{\sqrt{72}} \left( \left[14,32, \frac{12-34}{\sqrt{2}} \right]
\left[56+78+90 \right] \right)$
\\ \hline
${\bf(3,1},{\frac{5}{3}}) +c.c.$
& $\widehat{\Phi}^{(3,1,\frac{5}{3})}_{(75, 0)}$
& $\frac{1}{\sqrt{24}} \left(13  \left[80,06,68 \right] \right)+c.c. $
\\
\hline
${\bf(3,3},{\frac{2}{3}}) +c.c.$
& $\widehat{\Phi}^{(3,3,\frac{2}{3})}_{(40, -4)}$
& $\frac{1}{\sqrt{24}} \left( \left[14,32, \frac{12-34}{\sqrt{2}} \right] \right)
\left[80,06,68 \right] +c.c.$
\\
\hline
${\bf(6,2},{-\frac{1}{6}}) +c.c.$
& $\widehat{\Phi}^{(6,2,-\frac{1}{6})}_{(40, -4)}$
& $\frac{1}{\sqrt{24}} \left(  \left[2,4 \right]
\left[580, 670, 689,
\frac{6 \left(90-78 \right)}{\sqrt{2}},
\frac{8 \left(56-90 \right)}{\sqrt{2}},
\frac{0 \left(78-56 \right)}{\sqrt{2}} \right] \right) +c.c.$
\\
\hline
${\bf(6,2},{\frac{5}{6}}) +c.c.$
& $\widehat{\Phi}^{(6,2,\frac{5}{6})}_{(75, 0)}$
& $\frac{1}{\sqrt{24}} \left(  \left[1,3 \right]
\left[580, 670, 689,
\frac{6 \left(90-78 \right)}{\sqrt{2}},
\frac{8 \left(56-90 \right)}{\sqrt{2}},
\frac{0 \left(78-56 \right)}{\sqrt{2}} \right] \right) +c.c.$
\\ \hline
${\bf(8,1},{0})$
& $\widehat{A}^{(8,1,0)}_{(24, 0)}$
& $\frac{i}{\sqrt{2}}
\left(58,50,70,76,96,98,\frac{1}{\sqrt{2}}\left[56-78 \right],
\frac{1}{\sqrt{6}}\left[56+78-{\it 2}[90] \right] \right) $
\\
& $\widehat{E}^{(8,1,0)}_{(24, 0)}$
& $\frac{1}{\sqrt{2}}
\left\{58,50,70,76,96,98,\frac{1}{\sqrt{2}}\left[56-78 \right],
\frac{1}{\sqrt{6}}\left[56+78-{\it 2}[90] \right] \right\} $
\\
& $\widehat{\Phi}^{(8,1,0)}_{(24, 0)}$
& $\frac{1}{\sqrt{72}}
\left( \left[ 5890,5078,7056,7690,9678,9856, \frac{1}{\sqrt{2}}\left[5690-7890 \right],
\right. \right. $\\
& &$ \left.
\frac{1}{\sqrt{6}} [{\it 2}[5678]-5690-7890] \right] $
\\
&& $  + \left[12+34 \right]
\left[58,50,70,76,96,98, \frac{1}{\sqrt{2}}\left[56-78 \right], \right. $
\\
&&$ \left.\left.
\frac{1}{\sqrt{6}}\left[56+78-{\it 2}[90] \right] \right] \right) $
\\
%
& $\widehat{\Phi}^{(8,1,0)}_{(75, 0)}$
& $\frac{1}{\sqrt{144}}
\left( {\it 2} \times \left[ 5890,5078,7056,7690,9678,9856, \frac{1}{\sqrt{2}}\left[5690-7890 \right],
\right. \right. $\\
& &$ \left.
\frac{1}{\sqrt{6}} [{\it 2}[5678]-5690-7890] \right] $
\\
&& $  - \left[12+34 \right]
\left[58,50,70,76,96,98, \frac{1}{\sqrt{2}}\left[56-78 \right], \right. $
\\
&&$ \left.\left.
\frac{1}{\sqrt{6}}\left[56+78-{\it 2}[90] \right] \right] \right) $
\\
\hline
${\bf(8,1},{1}) +c.c.$
& $\widehat{\Phi}^{(8,1,1)}_{(\overline{40}, 4)}$
& $\frac{1}{\sqrt{24}} \left(13
\left[58,50,70,76,96,98, \frac{1}{\sqrt{2}}\left[56-78 \right], \right.\right. $
\\
&&$ \left.\left.
\frac{1}{\sqrt{6}}\left[56+78-{\it 2}[90] \right] \right] \right) +c.c. $
\\
\hline
${\bf(8,3},{0})$
& $\widehat{\Phi}^{(8,3,0)}_{(75, 0)}$
& $\frac{1}{\sqrt{24}} \left( \left[14,32, \frac{12-34}{\sqrt{2}} \right]
\left[58,50,70,76,96,98, \frac{1}{\sqrt{2}}\left[56-78 \right], \right.\right. $
\\
&&$ \left.\left.
\frac{1}{\sqrt{6}}\left[56+78 - {\it 2} [90] \right] \right] \right)$
\\
\hline \hline
\end{tabular}
\end{center}
\end{table}
\clearpage

\section{Examples of the CGC calculations}

Using the states in (\ref{vev}) and in Table \ref{tabw}-\ref{tabo2}
all the CGCs can be calculated directly.
For example, in the coupling\cite{sato,fuku}
\begin{eqnarray}
	-iA \overline{\Delta} \Delta
	&\equiv& -iA_{a'b'} \overline{\Delta}_{a'c'd'e'f'} \Delta_{b'c'd'e'f'}
	\ =\ -iA_{\overline{a}\overline{b}} \overline{\Delta}_{acdef}
	\Delta_{b\overline{c}\overline{d}\overline{e}\overline{f}},
\end{eqnarray}
we have
\begin{eqnarray}
	(-i)(a_1\widehat{a_1}) \cdot (\overline{V_R} \,\widehat{\overline{V_R}}) \cdot  (V_R\widehat{V_R})
	 &\equiv&
	 a_1 V_R \overline{V_R} (-i)(\widehat{a_1} \cdot \widehat{\overline{V_R}} \cdot \widehat{V_R} ) \nonumber \\
	 &=& a_1 V_R \overline{V_R} \frac{1}{\sqrt{10}} \frac{1}{\sqrt{120}} \frac{1}{\sqrt{120}}
		  (12 + 34 + 56 + 78 + 90) \cdot (13579) \cdot (24680) \nonumber \\
	 &=& a_1 V_R \overline{V_R}  \frac{1}{120\sqrt{10}} \times
	 	 5  \times  (-21) \cdot (13579) \cdot (24680) \nonumber \\
	 &=& a_1 V_R \overline{V_R}  \frac{1}{120\sqrt{10}} \times 5 \times (-1) \times 4! \nonumber \\
	 &=& a_1 V_R \overline{V_R} (-\frac{1}{\sqrt{10}}).
\end{eqnarray}
For another example, in the coupling
\begin{equation}
	\frac{1}{120} \varepsilon A \Phi^2
=\frac{1}{120}
	(-i)\varepsilon_{\overline{a}_1 \overline{a}_2 \overline{a}_3 \overline{a}_4 \overline{a}_5
		\overline{a}_6 \overline{a}_7 \overline{a}_8 \overline{a}_9 \overline{a}_0}
	A_{a_1 a_2} \Phi_{a_3 a_4 a_5 a_6} \Phi_{a_7 a_8 a_9 a_0},
\end{equation}
we have
\begin{eqnarray}
	&&\frac{1}{120}\varepsilon (a_2\widehat{a_2}) \cdot
	(\Phi_{(\overline{5}, 8)}^{(1, 2, -\frac{1}{2})} \widehat{\Phi}_{(\overline{5}, 8)}^{(1, 2, -\frac{1}{2})}) \cdot
	(\Phi_{(5, -8)}^{(1,2,\frac{1}{2})} \widehat{\Phi}_{(5, -8)}^{(1,2,\frac{1}{2})}) \nonumber \\
	&=& a_2 \Phi_{(\overline{5}, 8)}^{(1, 2, -\frac{1}{2})} \Phi_{(5, -8)}^{(1,2,\frac{1}{2})}
	(\frac{1}{120}\varepsilon \cdot \widehat{a_2} \cdot \widehat{\Phi}_{(\overline{5}, 8)}^{(1, 2, -\frac{1}{2})} \cdot \widehat{\Phi}_{(5, -8)}^{(1,2,\frac{1}{2})}) \nonumber \\
	&=& a_2 \Phi_{(\overline{5}, 8)}^{(1, 2, -\frac{1}{2})} \Phi_{(5, -8)}^{(1,2,\frac{1}{2})}
 \frac{1}{120} \times \frac{i}{\sqrt{60}} \times (\frac{1}{\sqrt{24}})^2
	\times (-i) (\overline{1} \overline{2} \overline{3} \overline{4} \overline{5}
	 \overline{6} \overline{7} \overline{8} \overline{9} \overline{0}) \nonumber \\
	&&\cdot 	 ({\it 3}\times[12 + 34] - {\it 2}\times[56 + 78 + 90])
	\cdot (-3579, 1579) \cdot (-4680, 2680)  \nonumber \\
	&=& a_2 \Phi_{(\overline{5}, 8)}^{(1, 2, -\frac{1}{2})} \Phi_{(5, -8)}^{(1,2,\frac{1}{2})}
 \frac{1}{24\times 120 \sqrt{60}} \times 3 \times   (\overline{1234567890})
	\cdot (12) \cdot (-3579)  \cdot (-4680) \nonumber \\
	&=& a_2 \Phi_{(\overline{5}, 8)}^{(1, 2, -\frac{1}{2})} \Phi_{(5, -8)}^{(1,2,\frac{1}{2})}
 \frac{1}{24\times 120\sqrt{60}} \times 3  \times 2! \times 4! \times 4! \nonumber \\
	&=& a_2 \Phi_{(\overline{5}, 8)}^{(1, 2, -\frac{1}{2})} \Phi_{(5, -8)}^{(1,2,\frac{1}{2})}
 \frac{1}{5} \sqrt{\frac{3}{5}}
\end{eqnarray}
There are two directions in  $\widehat{\Phi}_{(\overline{5}, 8)}^{(1, 2, -\frac{1}{2})}$
but only one of which needs to be counted in the calculations.

All the CGCs can be calculated following the examples given above.
In the following, we will focus on the couplings with at least one field with large VEV,
since they are relevant to the symmetry breaking and the masses.
Only the CGCs with at least one SM singlet will be given,
although the other CGCs can be calculated  without problems.

\section{The superpotential and the symmetry breaking}

The most general renormalizable Higgs superpotential is \cite{fuku}
\bea
W &=& \frac{1}{2} m_{1} \Phi^2 + m_{2} \overline{\Delta} \Delta + \frac{1}{2} m_{3} H^2+ \frac{1}{2} m_{4} A^2 + \frac{1}{2} m_{5} E^2 + \frac{1}{2} m_{6} D^2
\nonumber\\
&+& \lambda_{1} \Phi^3 + \lambda_{2} \Phi \overline{\Delta} \Delta
+ \left(\lambda_3 \Delta + \lambda_4 \overline{\Delta} \right) H \Phi
+\lambda_{5} A^2 \Phi -i \lambda_{6} A \overline{\Delta} \Delta
+ \frac{\lambda_7}{120} \varepsilon A \Phi^2
\nonumber\\
&+& E \left( \lambda_{8} E^2 + \lambda_{9} A^2 + \lambda_{10} \Phi^2
+ \lambda_{11} \Delta^2 + \lambda_{12} \overline{\Delta}^2 + \lambda_{13} H^2
\right)
+D^2 \left( \lambda_{14} E + \lambda_{15} \Phi \right)
\nonumber\\
&+& D \left\{ \lambda_{16} H A + \lambda_{17} H \Phi + \left(
\lambda_{18} \Delta + \lambda_{19} \overline{\Delta} \right) A
+ \left( \lambda_{20} \Delta + \lambda_{21} \overline{\Delta} \right) \Phi
\right\},
\label{potential}
\eea
only the SM singlets are relevant to the symmetry breaking into the SM gauge group.
Inserting the VEVs into ($\ref{potential}$), one obtains
\begin{eqnarray}
	\left<W\right> &=& \frac{1}{2} m_1 \left[ \phi_1^2 + \phi_2^2 + \phi_3^2 \right] + m_2 V_R \overline{V_R}
+ \frac{1}{2}m_4 (a_1^2 + a_2^2)
	+ \frac{1}{2}m_5 E^2 \nonumber \\
	&+& \lambda_1 \Big[ \phi_1^3 \frac{1}{2 \sqrt{15}}
	- \phi_2^3\frac{7}{54\sqrt{15}}
	+  \phi_3^3\frac{4}{27\sqrt{3}} \nonumber \\
	&&
	+ 3 \phi_1(\phi_2^2 - 2\phi_3^2) \frac{1}{12\sqrt{15}}
	- 3 \phi_2\phi_3^2\frac{4}{27\sqrt{15}}
	+3 \phi_3 \phi_2^2 \frac{5}{108\sqrt{3}}\Big] \nonumber \\
	&+& \lambda_2\phi_1 V_R \overline{V_R} \frac{1}{2\sqrt{15}}\nonumber \\
	&+& \lambda_5\left[a_1^2\phi_1 \frac{2}{\sqrt{15}}
	+ a_2^2\phi_1 (-\frac{1}{2\sqrt{15}})
	- a_2^2 \phi_2 \frac{1}{3\sqrt{15}}
	+ a_2^2\phi_3 \frac{5}{6\sqrt{3}}
	+ 2 a_1 a_2 \phi_2 \frac{1}{\sqrt{10}}\right] \nonumber \\
	&+& \lambda_6 a_1 V_R \overline{V_R} (-\frac{1}{\sqrt{10}})\nonumber \\
	&+& \lambda_7 \Bigg[ a_1\phi_1^2 \frac{6}{5\sqrt{10}} - a_1\phi_2^2 \frac{4}{5\sqrt{10}}+ a_1\phi_3^2\frac{2}{5\sqrt{10}}   \nonumber \\
	&&   - 2 a_2 \phi_1 \phi_2\frac{3}{5\sqrt{15}} + 2 a_2 \phi_2\phi_3 \frac{1}{3\sqrt{3}} + a_2\phi_2^2 \frac{4}{15\sqrt{15}}  - a_2\phi_3^2\frac{16}{15\sqrt{15}} \Bigg] \nonumber \\
	&+& \lambda_8E^3 \frac{1}{2\sqrt{15}} + \lambda_9E \left[a_1a_2 \frac{2}{\sqrt{10}}  + a_2^2 \frac{1}{2\sqrt{15}}\right ]  \nonumber \\
	&+& \lambda_{10} E \left [\phi_2^2 \frac{1}{12 \sqrt{15}} + \phi_3^2 \frac{2}{3\sqrt{15}} + 2 \phi_1\phi_2\frac{3}{4\sqrt{15}} + 2 \phi_2\phi_3 \frac{5}{12\sqrt{3}} \right].
	\label{VEVsCG}
\end{eqnarray}
Keeping SUSY at the high energy scale  requires
the D-flatness condition which is
    \begin{eqnarray}
        |V_R| = |\overline{V_R}|.
   \end{eqnarray}
 These VEVs are determined by the F-flatness conditions required by SUSY,
\begin{equation}
	\left\{
	\frac{\partial}{\partial{\phi_1}}, \,
	\frac{\partial}{\partial{\phi_2}}, \,
	\frac{\partial}{\partial{\phi_3}}, \,
	\frac{\partial}{\partial{V_R}}, \,
	\frac{\partial}{\partial{\overline{V_R}}}, \,
	\frac{\partial}{\partial{a_1}}, \,
	\frac{\partial}{\partial{a_2}}, \,
	\frac{\partial}{\partial{E}}
	\right\} \left< W \right> =0,
	\label{VEVs}
\end{equation}
from which we have
\begin{eqnarray}
    	0 &=&
    	m_1\phi_1
    	+\lambda_1 \phi_1^2 \frac{3}{2\sqrt{15}}
    	+\lambda_1 (\phi_2^2 - 2\phi_3^2) \frac{1}{4\sqrt{15}}
    	+\lambda_2 V_R \overline{V_R} \frac{1}{2\sqrt{15}}
    	+\lambda_5 a_1^2 \frac{2}{\sqrt{15}}
    	- \lambda_5a_2^2 \frac{1}{2\sqrt{15}} \nonumber \\
    	&&
    	+ \lambda_7 a_1 \phi_1 \frac{12}{5\sqrt{10}}
    	- \lambda_7 a_2 \phi_2  \frac{6}{5\sqrt{15}}
    	+ \lambda_{10} E \phi_2  \frac{3}{2\sqrt{15}},
    	\nonumber \\
    	0 &=&
    	m_1 \phi_2
    	-\lambda_1 \phi_2^2 \frac{7}{18\sqrt{15}}
    	+\lambda_1\phi_1\phi_2 \frac{1}{2\sqrt{15}}
    	-\lambda_1 \phi_3^2 \frac{4}{9\sqrt{15}}
    	+\lambda_1 \phi_2\phi_3 \frac{5}{18\sqrt{3}}
    	-\lambda_5 a_2^2 \frac{1}{3\sqrt{15}}
    	+\lambda_5 a_1 a_2 \frac{2}{\sqrt{10}}\nonumber \\
    	&&
    	-\lambda_7 a_1 \phi_2 \frac{8}{5\sqrt{10}}
    	-\lambda_7 a_2 \phi_1 \frac{6}{5\sqrt{15}}
    	+ \lambda_7 a_2 \phi_3 \frac{2}{3\sqrt{3}}
    	+ \lambda_7 a_2 \phi_2 \frac{8}{15\sqrt{15}} \nonumber \\
    	&&
    	+ \lambda_{10} E \phi_2 \frac{1}{6\sqrt{15}}
    	+ \lambda_{10} E \phi_1 \frac{3}{2\sqrt{15}}
    	+ \lambda_{10} \phi_3 \frac{5}{6\sqrt{3}}
    	\nonumber \\
    	0 &=&
    	m_1 \phi_3
    	+\lambda_1 \phi_3^2 \frac{4}{9\sqrt{3}}
    	-\lambda_1 \phi_1 \phi_3 \frac{1}{\sqrt{15}}
    	- \lambda_1 \phi_2\phi_3 \frac{8}{9\sqrt{15}}
    	+ \lambda_1 \phi_2^2 \frac{5}{36\sqrt{3}}
    	+ \lambda_5 a_2^2 \frac{5}{6\sqrt{3}}
    	\nonumber \\
    	&&
    	+ \lambda_7 a_1 \phi_3 \frac{4}{5\sqrt{10}}
    	+ \lambda_7 a_2 \phi_2 \frac{2}{3\sqrt{3}}
    	- \lambda_7 a_2 \phi_3 \frac{32}{15\sqrt{15}}
    	+ \lambda_{10} E \phi_3 \frac{4}{3\sqrt{15}}
    	+ \lambda_{10} E \phi_2 \frac{5}{6\sqrt{3}}
    	,
    	\nonumber \\
    	0 &=&
    	V_R ~{\rm or}~ \overline{V_R}
    	\left[
    	m_2
    	+\lambda_2 \phi_1 \frac{1}{2\sqrt{15}}
    	-\lambda_6 a_1 \frac{1}{\sqrt{10}}
    	\right],
    	\nonumber \\
    	0 &=&
    	m_4 a_1
    	+\lambda_5 a_1 \phi_1 \frac{4}{\sqrt{15}}
    	+ \lambda_5 a_2 \phi_2 \frac{2}{\sqrt{10}}
        - \lambda_6 V_R \overline{V_R}{\frac{1}{\sqrt{10}}}
    	+\lambda_7 \phi_1^2  \frac{6}{5\sqrt{10}} \nonumber \\
        &&
    	- \lambda_7  \phi_2^2 \frac{4}{5\sqrt{10}}
    	+\lambda_7 \phi_3^2  \frac{2}{5\sqrt{10}}
    	+ \lambda_9 E a_2 \frac{2}{\sqrt{10}} ,
    	\nonumber \\
    	0 &=&
    	m_4 a_2
        + \lambda_5 a_1 \phi_2 \frac{2}{\sqrt{10}}
        -\lambda_5 a_2 \phi_1 \frac{1}{\sqrt{15}}
    	- \lambda_5 a_2 \phi_2 \frac{2}{3\sqrt{15}}
    	+ \lambda_5 a_2 \phi_3 \frac{5}{3\sqrt{3}}
    	- \lambda_7 \phi_1 \phi_2 \frac{6}{5\sqrt{15}}
    	\nonumber \\
    	&&
        + \lambda_7  \phi_2 \phi_3 \frac{2}{3\sqrt{3}}
        + \lambda_7 \phi_2^2  \frac{4}{15\sqrt{15}}
    	- \lambda_7 \phi_3^2 \frac{16}{15\sqrt{15}}
        + \lambda_9 E a_1 \frac{2}{\sqrt{10}}
    	+ \lambda_9 E a_2 \frac{1}{\sqrt{15}},
    	\nonumber \\
    	0 &=&
    	m_5 E
    	+\lambda_8 E^2 \frac{3}{2\sqrt{15}}
    	+\lambda_9 a_1a_2\frac{2}{\sqrt{10}}
    	+ \lambda_9 a_2^2 \frac{1}{2\sqrt{15}} \nonumber \\
    	&&
    	+\lambda_{10}  \phi_2^2 \frac{1}{12 \sqrt{15}}
    	+ \lambda_{10} \phi_3^2 \frac{2}{3\sqrt{15}}
    	+ \lambda_{10}  \phi_1\phi_2\frac{3}{2\sqrt{15}}
    	+ \lambda_{10}  \phi_2\phi_3 \frac{5}{6\sqrt{3}}
    	\label{VEVeqs}.
\end{eqnarray}

\section{Masses and mass matrices}

The masses and mass matrices
for the  many states in the general model can be given once the VEVs are determined
by the superpotential parameters of the model.
They are presented in the form
$$M_{R_{(I,x)}^i S_{(J,y)}^j}=M_R\delta_{R\overline S} \delta_{I\overline J}\delta_{x,-y}+\sum_{T_K} \lambda_{RST}
C_{R_{(I,x)}^i S_{(J,y)}^j T_{(K,-x-y)}^{(1,1,0)}}<T_K>$$,
where $\{\lambda_{RST}\}\ =\ $
$\{6 \lambda_{1}$, $6 \lambda_{8}$, $2 \lambda_{5}$, $2 \lambda_{7}$, $2 \lambda_{9}$,
$2 \lambda_{10}$,  $2 \lambda_{11}$,  $2 \lambda_{12}$,  $2 \lambda_{13}$,
$2 \lambda_{14}$ , $2 \lambda_{15}$,
$\lambda_{2}$, $\lambda_{3}$, $\lambda_{4}$,
$\lambda_{6}$, $\lambda_{16}$, $\lambda_{17}$, $\lambda_{18}$,
$\lambda_{19}$, $\lambda_{20}$, $\lambda_{21}\}$, and
$C_{R_{(I,x)}^i S_{(J,y)}^j T_{(K,-x-y)}^{(1,1,0)}}$ is a CGC calculated in the $G_{51}$ basis
which is preformed in the present study.
We give the masses and mass matrices in the following,
from which the CGCs calculated in the present work in the $G_{51}$ basis can be read off.

%
%
\noindent
\begin{minipage}{16cm}
	$({\bf 1,1,}0)$\\[.2cm]
	{\bf c:}
	$\widehat{A}^{(1,1,0)}_{(1, 0)}$,
	$\widehat{A}^{(1,1,0)}_{(24, 0)}$,
	$\widehat{E}^{(1,1,0)}_{(24, 0)}$,
	$\widehat{\Delta}^{(1,1,0)}_{(1, -10)}$,
	$\widehat{\overline{\Delta}}^{(1,1,0)}_{(1, 10)}$,
	$\widehat{\Phi}^{(1,1,0)}_{(1, 0)}$,
	$\widehat{\Phi}^{(1,1,0)}_{(24, 0)}$,
	$\widehat{\Phi}^{(1,1,0)}_{(75, 0)}$\\[.15cm]
	{\bf r:}
	$\widehat{A}^{(1,1,0)}_{(1, 0)}$,
	$\widehat{A}^{(1,1,0)}_{(24, 0)}$,
	$\widehat{E}^{(1,1,0)}_{(75, 0)}$,
	$\widehat{\overline{\Delta}}^{(1,1,0)}_{(1, 10)}$,
	$\widehat{\Delta}^{(1,1,0)}_{(1, -10)}$,
	$\widehat{\Phi}^{(1,1,0)}_{(1, 0)}$,
	$\widehat{\Phi}^{(1,1,0)}_{(24, 0)}$,
	$\widehat{\Phi}^{(1,1,0)}_{(75, 0)}$\\[.2cm]
	\begin{equation}
	\left(
	\begin{array}{cccccccc}
	m_{11}^{(1,1,0)} &
	m_{12}^{(1,1,0)} &
	\sqrt{\frac{2}{5}}	a_2 \lambda _9 &
	-\frac{\overline{V_R} \lambda _6}{\sqrt{10}} &
	-\frac{V_R \lambda _6}{\sqrt{10}} &
	m_{16}^{(1,1,0)} &
	m_{17}^{(1,1,0)} &
	\frac{2}{5} \sqrt{\frac{2}{5}}\lambda _7 \phi _3
	\\
	m_{12}^{(1,1,0)} &
	m_{22}^{(1,1,0)} &
	m_{23}^{(1,1,0)} &
	0 &
	0 &
	m_{26}^{(1,1,0)} &
	m_{27}^{(1,1,0)} &
	m_{28}^{(1,1,0)}
	\\
	\sqrt{\frac{2}{5}} a_2 \lambda _9 &
	m_{23}^{(1,1,0)} &
	m_{33}^{(1,1,0)} &
	0 &
	0 &
	\frac{1}{2} \sqrt{\frac{3}{5}} \lambda _{10} \phi _2 &
	m_{37}^{(1,1,0)} &
	m_{38}^{(1,1,0)}
	\\
	-\frac{V_R \lambda _6}{\sqrt{10}} &
	0 &
	0 &
	m_{44}^{(1,1,0)} &
	0 &
	\frac{V_R \lambda_2}{2 \sqrt{15}} &
	0 &
	0
	\\
	-\frac{\overline{V_R} \lambda _6}{\sqrt{10}} &
	0 &
	0 &
	0 &
	m_{44}^{(1,1,0)} &
	\frac{\overline{V_R}\lambda _2}{2 \sqrt{15}} &
	0 &
	0
	\\
	m_{16}^{(1,1,0)} &
	m_{26}^{(1,1,0)} &
	\frac{1}{2} \sqrt{\frac{3}{5}} \lambda _{10} \phi _2 &
	\frac{\overline{V_R} \lambda _2}{2 \sqrt{15}} & \frac{V_R \lambda _2}{2\sqrt{15}} &
	m_{66}^{(1,1,0)} &
	m_{67}^{(1,1,0)} &
	-\frac{\lambda _1 \phi _3}{\sqrt{15}}
	\\
	m_{17}^{(1,1,0)}&
	m_{27}^{(1,1,0)} &
	m_{37}^{(1,1,0)} &
	0 &
	0 &
	m_{67}^{(1,1,0)} &
	m_{77}^{(1,1,0)} &
	m_{78}^{(1,1,0)}
	\\
	\frac{2}{5} \sqrt{\frac{2}{5}} \lambda _7 \phi _3 &
	m_{28}^{(1,1,0)} &
	m_{38}^{(1,1,0)} &
	0 &
	0 &
	-\frac{\lambda _1 \phi_3}{\sqrt{15}} &
	m_{78}^{(1,1,0)} &
	m_{88}^{(1,1,0)}
	\\
	\end{array}
	\right),
	\end{equation}
\end{minipage}

\noindent  where
\begin{eqnarray}
m_{11}^{(1,1,0)} &\equiv&
m_4 + \frac{4 \lambda _5 \phi _1}{\sqrt{15}},
\nonumber\\
m_{12}^{(1,1,0)} &\equiv&
\sqrt{\frac{2}{5}} E \lambda _9+\sqrt{\frac{2}{5}} \lambda _5 \phi _2,
\nonumber \\
m_{16}^{(1,1,0)} &\equiv&
\frac{4 a_1 \lambda _5}{\sqrt{15}}+\frac{6}{5} \sqrt{\frac{2}{5}} \lambda _7 \phi _1,
\nonumber\\
m_{17}^{(1,1,0)} &\equiv&
\sqrt{\frac{2}{5}} a_2 \lambda _5-\frac{4}{5} \sqrt{\frac{2}{5}} \lambda _7 \phi _2,
\nonumber\\
m_{22}^{(1,1,0)} &\equiv&
m_4+\frac{E \lambda _9}{\sqrt{15}}-\frac{\lambda _5 \phi_1}{\sqrt{15}}-\frac{2 \lambda _5 \phi _2}{3 \sqrt{15}}+\frac{5 \lambda _5 \phi _3}{3 \sqrt{3}},
\nonumber\\
m_{23}^{(1,1,0)} &\equiv&
\sqrt{\frac{2}{5}} a_1 \lambda _9+\frac{a_2 \lambda_9}{\sqrt{15}},
\nonumber\\
m_{26}^{(1,1,0)} &\equiv&
-\frac{a_2 \lambda _5}{\sqrt{15}}-\frac{2}{5} \sqrt{\frac{3}{5}}
\lambda _7 \phi _2,
\nonumber\\
m_{27}^{(1,1,0)} &\equiv&
\sqrt{\frac{2}{5}} a_1 \lambda _5-\frac{2 a_2 \lambda _5}{3
	\sqrt{15}}-\frac{2}{5} \sqrt{\frac{3}{5}} \lambda _7 \phi _1+\frac{8 \lambda _7 \phi _2}{15 \sqrt{15}}+\frac{2 \lambda _7 \phi _3}{3 \sqrt{3}},
\nonumber\\
m_{28}^{(1,1,0)} &\equiv&
\frac{5 a_2 \lambda _5}{3 \sqrt{3}}+\frac{2 \lambda _7 \phi _2}{3 \sqrt{3}}-\frac{32 \lambda
	_7 \phi _3}{15 \sqrt{15}},
\nonumber\\
m_{33}^{(1,1,0)} &\equiv&
m_5+\sqrt{\frac{3}{5}} E \lambda_8,
\nonumber\\
m_{37}^{(1,1,0)} &\equiv&
\frac{1}{2} \sqrt{\frac{3}{5}} \lambda _{10} \phi _1+\frac{\lambda _{10} \phi _2}{6 \sqrt{15}}+\frac{5 \lambda _{10} \phi _3}{6 \sqrt{3}},
\nonumber \\
m_{38}^{(1,1,0)} &\equiv&
\frac{5 \lambda _{10} \phi _2}{6 \sqrt{3}}+\frac{4 \lambda _{10} \phi _3}{3 \sqrt{15}},
\nonumber\\
m_{44}^{(1,1,0)} &\equiv&
m_2-\frac{a_1 \lambda _6}{\sqrt{10}}+\frac{\lambda _2 \phi _1}{2 \sqrt{15}},
\nonumber\\
m_{66}^{(1,1,0)} &\equiv&
m_1+\frac{6}{5} \sqrt{\frac{2}{5}} a_1 \lambda _7+\sqrt{\frac{3}{5}} \lambda _1 \phi _1,
\nonumber\\
m_{67}^{(1,1,0)} &\equiv&
-\frac{2}{5} \sqrt{\frac{3}{5}} a_2 \lambda _7+\frac{1}{2} \sqrt{\frac{3}{5}} E \lambda _{10}+\frac{\lambda _1 \phi _2}{2 \sqrt{15}},
\nonumber\\
m_{77}^{(1,1,0)} &\equiv&
m_1-\frac{4}{5} \sqrt{\frac{2}{5}} a_1 \lambda _7+\frac{8 a_2 \lambda _7}{15 \sqrt{15}}+\frac{E \lambda _{10}}{6 \sqrt{15}}+\frac{\lambda
	_1 \phi _1}{2 \sqrt{15}}-\frac{7 \lambda _1 \phi _2}{9 \sqrt{15}}+\frac{5 \lambda _1 \phi _3}{18 \sqrt{3}},
\nonumber\\
m_{78}^{(1,1,0)} &\equiv&
\frac{2 a_2 \lambda _7}{3 \sqrt{3}}+\frac{5 E \lambda _{10}}{6 \sqrt{3}}+\frac{5 \lambda _1 \phi _2}{18\sqrt{3}}-\frac{8
	\lambda _1 \phi _3}{9 \sqrt{15}},
\nonumber \\
m_{88}^{(1,1,0)} &\equiv&
m_1+\frac{2}{5} \sqrt{\frac{2}{5}} a_1 \lambda _7-\frac{32 a_2 \lambda _7}{15 \sqrt{15}}+\frac{4 E \lambda_{10}}{3 \sqrt{15}}-\frac{\lambda _1 \phi _1}{\sqrt{15}}-\frac{8 \lambda _1 \phi _2}{9 \sqrt{15}}+\frac{8 \lambda _1 \phi _3}{9 \sqrt{3}}.
\nonumber \\
\end{eqnarray}\\[.2cm]
\begin{minipage}{16cm}
	$\left[{\bf{(1,1}},1) +c.c. \right]$\\[.2cm]
	{\bf c:}
	$\widehat{A}^{(1,1,1)}_{(10, 4)}$,
	$\widehat{D}^{(1,1,1)}_{(10, -6)}$,
	$\widehat{\Delta}^{(1,1,1)}_{(10, -6)}$
	$\widehat{\Phi}^{(1,1,1)}_{(10, 4)}$ \\[.15cm]
	{\bf r:}
	$\widehat{A}^{(1,1,-1)}_{(\overline{10}, -4)}$,
	$\widehat{D}^{(1,1,-1)}_{(\overline{10}, 6)}$,
	$\widehat{\overline{\Delta}}^{(1,1,-1)}_{(\overline{10}, 6)}$
	$\widehat{\Phi}^{(1,1,-1)}_{(\overline{10}, -4)}$ \\[.2cm]
	
	\begin{equation}
	\left(
	\begin{array}{cccc}
	m_{11}^{(1,1,1)}&
	-\frac{i \overline{V_R} \lambda _{19}}{\sqrt{10}} &
	-\frac{1}{5} \overline{V_R} \lambda _6 &
	m_{14}^{(1,1,1)}
	\\
	\frac{i V_R \lambda _{18}}{\sqrt{10}} &
	m_{22}^{(1,1,1)} &
	m_{23}^{(1,1,1)} &
	-\frac{V_R \lambda_{20}}{2\sqrt{10}}
	\\
	-\frac{1}{5} V_R \lambda _6 &
	m_{32}^{(1,1,1)}&
	m_{33}^{(1,1,1)}&
	-\frac{1}{10} V_R \lambda _2
	\\
	m_{14}^{(1,1,1)} &
	-\frac{\overline{V_R} \lambda _{21}}{2\sqrt{10}} &
	-\frac{1}{10} \overline{V_R} \lambda_2 &
	m_{44}^{(1,1,1)}
	\\
	\end{array}
	\right),
	\end{equation}
\end{minipage}

\noindent where
\begin{eqnarray}
m_{11}^{(1,1,1)} &\equiv&
m_4+\sqrt{\frac{3}{5}} E \lambda _9+\frac{\lambda _5 \phi _1}{\sqrt{15}}+\frac{2 \lambda _5 \phi _2}{\sqrt{15}}+\frac{\lambda _5 \phi _3}{\sqrt{3}}, \nonumber\\
m_{14}^{(1,1,1)} &\equiv&
-\sqrt{\frac{2}{5}} a_1 \lambda _5+\frac{2 a_2 \lambda
	_5}{\sqrt{15}}-\frac{2}{5} \sqrt{\frac{3}{5}} \lambda _7 \phi _1+\frac{8 \lambda _7 \phi _2}{5 \sqrt{15}}-\frac{2 \lambda _7 \phi _3}{5 \sqrt{3}}, \nonumber \\
m_{22}^{(1,1,1)} &\equiv&
m_6-\frac{2 E \lambda _{14}}{\sqrt{15}}+\frac{\lambda _{15} \phi _1}{\sqrt{15}}-\frac{4 \lambda _{15} \phi
	_2}{3 \sqrt{15}}+\frac{\lambda _{15} \phi _3}{3 \sqrt{3}}, \nonumber \\
m_{23}^{(1,1,1)} &\equiv&
-\frac{1}{5} i a_1 \lambda _{18}-\frac{1}{5} i \sqrt{\frac{3}{2}} a_2 \lambda _{18}-\frac{1}{10}
\sqrt{\frac{3}{2}} \lambda _{20} \phi _1-\frac{\lambda _{20} \phi _2}{10 \sqrt{6}}+\frac{\lambda _{20} \phi _3}{2 \sqrt{30}}, \nonumber \\
m_{32}^{(1,1,1)} &\equiv&
\frac{1}{5} i a_1 \lambda _{19}+\frac{1}{5} i \sqrt{\frac{3}{2}} a_2 \lambda _{19}-\frac{1}{10} \sqrt{\frac{3}{2}}
\lambda _{21} \phi _1-\frac{\lambda _{21} \phi _2}{10 \sqrt{6}}+\frac{\lambda _{21} \phi _3}{2 \sqrt{30}}, \nonumber \\
m_{33}^{(1,1,1)} &\equiv&
m_2-\frac{3 a_1 \lambda _6}{5 \sqrt{10}}+\frac{1}{5}\sqrt{\frac{3}{5}} a_2 \lambda _6+\frac{\lambda _2 \phi _1}{5 \sqrt{15}}-\frac{\lambda _2 \phi _2}{10 \sqrt{15}}+\frac{\lambda _2 \phi _3}{10 \sqrt{3}}, \nonumber \\
m_{44}^{(1,1,1)} &\equiv&
m_1+\frac{4}{5} \sqrt{\frac{2}{5}} a_1 \lambda _7-\frac{8 a_2 \lambda _7}{5 \sqrt{15}}+\frac{E \lambda _{10}}{2 \sqrt{15}}+\frac{1}{2} \sqrt{\frac{3}{5}}
\lambda _1 \phi _1-\frac{\lambda _1 \phi _2}{3 \sqrt{15}}+\frac{5 \lambda _1 \phi _3}{6 \sqrt{3}}.
\end{eqnarray}\\[.2cm]
\begin{minipage}{16cm}
	$\left[{\bf{(3,1}}, \frac{2}{3}) +c.c. \right]$\\[.2cm]
	{\bf c:}
	$\widehat{A}^{(3,1,\frac{2}{3})}_{(\overline{10}, -4)}$,
	$\widehat{D}^{(3,1,\frac{2}{3})}_{(\overline{10}, 6)}$,
	$\widehat{\overline{\Delta}}^{(3,1,\frac{2}{3})}_{(\overline{10}, 6)}$,
	$\widehat{\Phi}^{(3,1,\frac{2}{3})}_{(\overline{10}, -4)}$,
	$\widehat{\Phi}^{(3,1,\frac{2}{3})}_{(40, -4)}$\\[.15cm]
	{\bf r:}
	$\widehat{A}^{(\overline{3},1,-\frac{2}{3})}_{(10, 4)}$,
	$\widehat{D}^{(\overline{3},1,-\frac{2}{3})}_{(10, -6)}$,
	$\widehat{\Delta}^{(\overline{3},1,-\frac{2}{3})}_{(10, -6)}$,
	$\widehat{\Phi}^{(\overline{3},1,-\frac{2}{3})}_{(10, 4)}$,
	$\widehat{\Phi}^{(\overline{3},1,-\frac{2}{3})}_{(\overline{40}, 4)}$\\
	\begin{equation}
	\left(
	\begin{array}{ccccc}
	m_{11}^{(3,1,\frac{2}{3})}&
	-\frac{i V_R \lambda _{18}}{\sqrt{10}} &
	-\frac{1}{5} V_R \lambda _6 &
	m_{14}^{(3,1,\frac{2}{3})}&
	m_{15}^{(3,1,\frac{2}{3})}
	\\
	\frac{i \overline{V_R} \lambda _{19}}{\sqrt{10}} &
	m_{22}^{(3,1,\frac{2}{3})} &
	m_{23}^{(3,1,\frac{2}{3})} &
	\frac{\overline{V_R} \lambda _{21}}{2 \sqrt{10}} &
	0
	\\
	-\frac{1}{5} \overline{V_R} \lambda _6 &
	m_{32}^{(3,1,\frac{2}{3})} &
	m_{33}^{(3,1,\frac{2}{3})} &
	-\frac{1}{10} \overline{V_R} \lambda _2 &
	0
	\\
	m_{14}^{(3,1,\frac{2}{3})} &
	\frac{V_R \lambda _{20}}{2 \sqrt{10}} &
	-\frac{1}{10} V_R \lambda _2 &
	m_{44}^{(3,1,\frac{2}{3})} &
	m_{45}^{(3,1,\frac{2}{3})}
	\\
	m_{15}^{(3,1,\frac{2}{3})} &
	0 &
	0 &
	m_{45}^{(3,1,\frac{2}{3})} &
	m_{55}^{(3,1,\frac{2}{3})}
	\\
	\end{array}
	\right),
	\end{equation}
\end{minipage}

\noindent where
\begin{eqnarray}
m_{11}^{(3,1,\frac{2}{3})} &\equiv&
m_4-\frac{2 E \lambda _9}{\sqrt{15}}+\frac{\lambda _5 \phi _1}{\sqrt{15}}-\frac{4 \lambda _5 \phi _2}{3 \sqrt{15}}+\frac{\lambda _5 \phi _3}{3 \sqrt{3}},
\nonumber\\
m_{14}^{(3,1,\frac{2}{3})} &\equiv&
-\sqrt{\frac{2}{5}} a_1 \lambda _5-\frac{4 a_2 \lambda _5}{3 \sqrt{15}}-\frac{2}{5}
\sqrt{\frac{3}{5}} \lambda _7 \phi _1-\frac{16 \lambda _7 \phi _2}{15 \sqrt{15}}-\frac{2 \lambda _7 \phi _3}{15 \sqrt{3}},
\nonumber \\
m_{15}^{(3,1,\frac{2}{3})} &\equiv&
\frac{1}{3} \sqrt{\frac{10}{3}} a_2 \lambda _5-\frac{2}{3} \sqrt{\frac{2}{15}} \lambda _7 \phi _2-\frac{8}{15} \sqrt{\frac{2}{3}} \lambda _7 \phi _3,
\nonumber \\
m_{22}^{(3,1,\frac{2}{3})} &\equiv&
m_6+\frac{4 E \lambda _{14}}{3 \sqrt{15}}+\frac{\lambda _{15} \phi _1}{\sqrt{15}}+\frac{8 \lambda
	_{15} \phi _2}{9 \sqrt{15}}+\frac{\lambda _{15} \phi _3}{9 \sqrt{3}},
\nonumber\\
m_{23}^{(3,1,\frac{2}{3})} &\equiv&
-\frac{1}{5} i a_1 \lambda _{19}+\frac{1}{5} i \sqrt{\frac{2}{3}} a_2 \lambda
_{19}+\frac{1}{10} \sqrt{\frac{3}{2}} \lambda _{21} \phi _1-\frac{\lambda _{21} \phi _2}{15 \sqrt{6}}-\frac{\lambda _{21} \phi _3}{6 \sqrt{30}},
\nonumber\\
m_{32}^{(3,1,\frac{2}{3})} &\equiv&
\frac{1}{5} i a_1 \lambda _{18}-\frac{1}{5} i \sqrt{\frac{2}{3}} a_2 \lambda _{18}+\frac{1}{10} \sqrt{\frac{3}{2}}
\lambda _{20} \phi _1-\frac{\lambda _{20} \phi _2}{15 \sqrt{6}}-\frac{\lambda _{20} \phi _3}{6 \sqrt{30}},
\nonumber\\
m_{33}^{(3,1,\frac{2}{3})} &\equiv&
m_2-\frac{3 a_1 \lambda _6}{5 \sqrt{10}}-\frac{2
	a_2 \lambda _6}{5 \sqrt{15}}+\frac{\lambda _2 \phi _1}{5 \sqrt{15}}+\frac{\lambda _2 \phi _2}{15 \sqrt{15}}+\frac{\lambda _2 \phi _3}{30 \sqrt{3}},
\nonumber\\
m_{44}^{(3,1,\frac{2}{3})} &\equiv&
m_1+\frac{4}{5}
\sqrt{\frac{2}{5}} a_1 \lambda _7+\frac{16 a_2 \lambda _7}{15 \sqrt{15}}-\frac{E \lambda _{10}}{3 \sqrt{15}}+\frac{1}{2} \sqrt{\frac{3}{5}} \lambda
_1 \phi _1+\frac{2 \lambda _1 \phi _2}{9 \sqrt{15}}+\frac{5 \lambda _1 \phi _3}{18 \sqrt{3}},
\nonumber\\
m_{45}^{(3,1,\frac{2}{3})} &\equiv&
\frac{2}{3} \sqrt{\frac{2}{15}} a_2 \lambda _7-\frac{1}{3}
\sqrt{\frac{5}{6}} E \lambda _{10}-\frac{1}{9} \sqrt{\frac{5}{6}} \lambda _1 \phi _2-\frac{2}{9} \sqrt{\frac{2}{3}} \lambda _1 \phi _3,
\nonumber\\
m_{55}^{(3,1,\frac{2}{3})} &\equiv&
m_1-\frac{2}{5} \sqrt{\frac{2}{5}} a_1 \lambda _7-\frac{28 a_2 \lambda _7}{15 \sqrt{15}}-\frac{7
	E \lambda _{10}}{6 \sqrt{15}}-\frac{1}{9} \sqrt{\frac{5}{3}} \lambda _1 \phi _2+\frac{5 \lambda _1 \phi _3}{9 \sqrt{3}}.
\end{eqnarray} \\[.2cm]
\begin{minipage}{16cm}
	$\left[{\bf{(3,2}}, -\frac{5}{6}) +c.c. \right]$\\[.2cm]
	{\bf c:}
	$\widehat{A}^{(3,2,-\frac{5}{6})}_{(24, 0)}$,
	$\widehat{E}^{(3,2,-\frac{5}{6})}_{(24, 0)}$,
	$\widehat{\Phi}^{(3,2,-\frac{5}{6})}_{(24, 0)}$
	$\widehat{\Phi}^{(3,2,-\frac{5}{6})}_{(75, 0)}$\\[.15cm]
	{\bf r:}
	$\widehat{A}^{(\overline{3},2,\frac{5}{6})}_{(24, 0)}$,
	$\widehat{E}^{(\overline{3},2,\frac{5}{6})}_{(24, 0)}$,
	$\widehat{\Phi}^{(\overline{3},2,\frac{5}{6})}_{(24, 0)}$
	$\widehat{\Phi}^{(\overline{3},2,\frac{5}{6})}_{(75, 0)}$\\[.2cm]
	\begin{equation}
	\left(
	\begin{array}{cccc}
	m_{11}^{(3,2,-\frac{5}{6})} &
	-\sqrt{\frac{2}{5}} a_1 \lambda _9-\frac{a_2 \lambda _9}{2 \sqrt{15}} &
	m_{13}^{(3,2,-\frac{5}{6})} &
	m_{14}^{(3,2,-\frac{5}{6})}
	\\
	-\sqrt{\frac{2}{5}} a_1 \lambda _9-\frac{a_2 \lambda _9}{2 \sqrt{15}} &
	\frac{1}{2} \sqrt{\frac{3}{5}} E+m_5 &
	m_{23}^{(3,2,-\frac{5}{6})} &
	\frac{1}{3} \sqrt{\frac{5}{6}} \lambda
	_{10} \phi _2+\frac{\lambda _{10} \phi _3}{3 \sqrt{6}}
	\\
	m_{13}^{(3,2,-\frac{5}{6})} &
	m_{23}^{(3,2,-\frac{5}{6})} &
	m_{33}^{(3,2,-\frac{5}{6})} &
	m_{34}^{(3,2,-\frac{5}{6})}
	\\
	m_{14}^{(3,2,-\frac{5}{6})} &
	\frac{1}{3} \sqrt{\frac{5}{6}} \lambda _{10} \phi _2+\frac{\lambda _{10} \phi _3}{3 \sqrt{6}} &
	m_{34}^{(3,2,-\frac{5}{6})} &
	m_{44}^{(3,2,-\frac{5}{6})}
	\\
	\end{array}
	\right),
	\end{equation}
\end{minipage}\\[.2cm]
where
\begin{eqnarray}
m_{11}^{(3,2,-\frac{5}{6})} &\equiv&
m_4+\frac{E \lambda _9}{2 \sqrt{15}}-\frac{\lambda _5 \phi _1}{\sqrt{15}}-\frac{\lambda _5 \phi _2}{3 \sqrt{15}}+\frac{\lambda _5 \phi _3}{3\sqrt{3}},
\nonumber \\
m_{13}^{(3,2,-\frac{5}{6})} &\equiv&
-\sqrt{\frac{2}{5}} a_1 \lambda _5+\frac{a_2 \lambda _5}{3 \sqrt{15}}+\frac{2}{5}
\sqrt{\frac{3}{5}} \lambda _7 \phi _1-\frac{4 \lambda _7 \phi _2}{15 \sqrt{15}}-\frac{2 \lambda _7 \phi _3}{15 \sqrt{3}},
\nonumber \\
m_{14}^{(3,2,-\frac{5}{6})} &\equiv&
-\frac{1}{3} \sqrt{\frac{10}{3}} a_2 \lambda _5-\frac{2}{3} \sqrt{\frac{2}{15}} \lambda _7 \phi _2+\frac{4}{15} \sqrt{\frac{2}{3}} \lambda _7 \phi_3,
\nonumber \\
m_{23}^{(3,2,-\frac{5}{6})} &\equiv&
\frac{1}{2} \sqrt{\frac{3}{5}}\lambda _{10} \phi _1+\frac{\lambda _{10} \phi _2}{12 \sqrt{15}}+\frac{\lambda _{10} \phi _3}{6 \sqrt{3}},
\nonumber \\
m_{33}^{(3,2,-\frac{5}{6})} &\equiv&
m_1-\frac{4}{5} \sqrt{\frac{2}{5}} a_1 \lambda _7+\frac{4 a_2 \lambda _7}{15 \sqrt{15}}+\frac{E
	\lambda _{10}}{12 \sqrt{15}}+\frac{\lambda _1 \phi _1}{2 \sqrt{15}}-\frac{7 \lambda _1 \phi _2}{18 \sqrt{15}}+\frac{\lambda _1 \phi _3}{18 \sqrt{3}},
\nonumber \\
m_{34}^{(3,2,-\frac{5}{6})} &\equiv&
\frac{2}{3} \sqrt{\frac{2}{15}} a_2 \lambda _7+\frac{1}{3} \sqrt{\frac{5}{6}} E \lambda _{10}+\frac{1}{9} \sqrt{\frac{5}{6}} \lambda _1
\phi _2-\frac{1}{9} \sqrt{\frac{2}{3}} \lambda _1 \phi _3,
\nonumber\\
m_{44}^{(3,2,-\frac{5}{6})} &\equiv&
m_1+\frac{2}{5} \sqrt{\frac{2}{5}} a_1 \lambda _7-\frac{22 a_2 \lambda _7}{15 \sqrt{15}}+\frac{11 E \lambda _{10}}{12 \sqrt{15}}-\frac{\lambda
	_1 \phi _1}{\sqrt{15}}-\frac{11 \lambda _1 \phi _2}{18 \sqrt{15}}+\frac{4 \lambda _1 \phi _3}{9 \sqrt{3}}.
\end{eqnarray} \\[0.2cm]
\begin{minipage}{16cm}
	$\left[{\bf{(3,2}}, \frac{1}{6}) +c.c. \right]$\\[.2cm]
	{\bf c:}
	$\widehat{A}^{(3,2,\frac{1}{6})}_{(10, 4)}$,
	$\widehat{E}^{(3,2,\frac{1}{6})}_{(15, 4)}$,
	$\widehat{D}^{(3,2,\frac{1}{6})}_{(10, -6)}$,
	$\widehat{\Delta}^{(3,2,\frac{1}{6})}_{(10, -6)}$,
	$\widehat{\overline{\Delta}}^{(3,2,\frac{1}{6})}_{(15, -6)}$,
	$\widehat{\Phi}^{(3,2,\frac{1}{6})}_{(10, 4)}$,
	$\widehat{\Phi}^{(3,2,\frac{1}{6})}_{(\overline{40}, 4)}$\\[.15cm]
	{\bf r:}
	$\widehat{A}^{(\overline{3},2,-\frac{1}{6})}_{(\overline{10}, -4)}$,
	$\widehat{E}^{(\overline{3},2,-\frac{1}{6})}_{(\overline{15}, -4)}$,
	$\widehat{D}^{(\overline{3},2,-\frac{1}{6})}_{(\overline{10}, 6)}$,
	$\widehat{\overline{\Delta}}^{(\overline{3},2,-\frac{1}{6})}_{(\overline{10}, 6)}$,
	$\widehat{\Delta}^{(\overline{3},2,-\frac{1}{6})}_{(\overline{15}, 6)}$,
	$\widehat{\Phi}^{(\overline{3},2,-\frac{1}{6})}_{(\overline{10}, -4)}$,
	$\widehat{\Phi}^{(\overline{3},2,-\frac{1}{6})}_{(40, -4)}$\\[.2cm]
	\begin{equation}
	\left(
	\begin{array}{ccccccc}
	m_{11}^{(3,2,\frac{1}{6})} &
	\frac{1}{2} \sqrt{\frac{5}{3}} a_2 \lambda _9 &
	-\frac{i \overline{V_R} \lambda _{19}}{\sqrt{10}} &
	-\frac{1}{5} \overline{V_R} \lambda_6 &
	0 &
	m_{16}^{(3,2,\frac{1}{6})} &
	m_{17}^{(3,2,\frac{1}{6})}
	\\
	\frac{1}{2} \sqrt{\frac{5}{3}} a_2 \lambda _9 &
	m_5+\frac{1}{2} \sqrt{\frac{3}{5}} E \lambda _8 &
	0 &
	0 &
	-\frac{2}{5} \overline{V_R} \lambda _{12} &
	-\frac{1}{4} \sqrt{\frac{5}{3}} \lambda _{10} \phi _2 &
	-\frac{\lambda _{10} \phi _3}{\sqrt{6}}
	\\
	\frac{i V_R \lambda _{18}}{\sqrt{10}} &
	0 &
	m_{33}^{(3,2,\frac{1}{6})} &
	m_{34}^{(3,2,\frac{1}{6})} &
	-\frac{i a_2 \lambda_{19}}{2 \sqrt{6}}-\frac{\lambda _{21} \phi _2}{4 \sqrt{6}} &
	-\frac{V_R \lambda _{20}}{2 \sqrt{10}} &
	0
	\\
	-\frac{1}{5} V_R \lambda _6 &
	0 &
	m_{43}^{(3,2,\frac{1}{6})}  &
	m_{44}^{(3,2,\frac{1}{6})}  &
	\frac{E \lambda _{12}}{\sqrt{15}} &
	-\frac{1}{10} V_R \lambda _2 &
	0
	\\
	0 &
	-\frac{2}{5} V_R \lambda _{11} &
	\frac{i a_2 \lambda _{18}}{2 \sqrt{6}}-\frac{\lambda _{20} \phi _2}{4 \sqrt{6}} &
	\frac{E \lambda _{11}}{\sqrt{15}}&
	m_{55}^{(3,2,\frac{1}{6})}  &
	0 &
	0
	\\
	m_{16}^{(3,2,\frac{1}{6})} &
	-\frac{1}{4} \sqrt{\frac{5}{3}} \lambda _{10} \phi _2 &
	-\frac{\overline{V_R} \lambda_{21}}{2 \sqrt{10}} &
	-\frac{1}{10} \overline{V_R} \lambda _2 &
	0 &
	m_{66}^{(3,2,\frac{1}{6})} &
	m_{67}^{(3,2,\frac{1}{6})}
	\\
	m_{17}^{(3,2,\frac{1}{6})} &
	-\frac{\lambda _{10} \phi _3}{\sqrt{6}} &
	0 &
	0 &
	0 &
	m_{67}^{(3,2,\frac{1}{6})} &
	m_{77}^{(3,2,\frac{1}{6})}
	\\
	\end{array}
	\right),
	\end{equation}
\end{minipage}\\[.2cm]
where
\begin{eqnarray}
m_{11}^{(3,2,\frac{1}{6})} &\equiv&
m_4+\frac{E \lambda _9}{2 \sqrt{15}}+\frac{\lambda _5 \phi _1}{\sqrt{15}}+\frac{\lambda _5 \phi _2}{3 \sqrt{15}}-\frac{\lambda _5 \phi _3}{3
	\sqrt{3}},
\nonumber \\
m_{16}^{(3,2,\frac{1}{6})} &\equiv&
-\sqrt{\frac{2}{5}} a_1 \lambda _5+\frac{a_2 \lambda _5}{3 \sqrt{15}}-\frac{2}{5} \sqrt{\frac{3}{5}} \lambda _7 \phi _1+\frac{4 \lambda _7 \phi
	_2}{15 \sqrt{15}}+\frac{2 \lambda _7 \phi _3}{15 \sqrt{3}},
\nonumber \\
m_{17}^{(3,2,\frac{1}{6})} &\equiv&
-\frac{1}{3} \sqrt{\frac{10}{3}} a_2 \lambda _5+\frac{2}{3} \sqrt{\frac{2}{15}}
\lambda _7 \phi _2-\frac{4}{15} \sqrt{\frac{2}{3}} \lambda _7 \phi _3,
\nonumber \\
m_{33}^{(3,2,\frac{1}{6})} &\equiv&
m_6-\frac{E \lambda _{14}}{3 \sqrt{15}}+\frac{\lambda _{15} \phi _1}{\sqrt{15}}-\frac{2 \lambda
	_{15} \phi _2}{9 \sqrt{15}}-\frac{\lambda _{15} \phi _3}{9 \sqrt{3}},
\nonumber\\
m_{34}^{(3,2,\frac{1}{6})} &\equiv&
-\frac{1}{5} i a_1 \lambda _{18}-\frac{i a_2 \lambda _{18}}{10 \sqrt{6}}-\frac{1}{10}
\sqrt{\frac{3}{2}} \lambda _{20} \phi _1-\frac{\lambda _{20} \phi _2}{60 \sqrt{6}}-\frac{\lambda _{20} \phi _3}{6 \sqrt{30}},
\nonumber\\
m_{43}^{(3,2,\frac{1}{6})} &\equiv&
\frac{1}{5} i a_1 \lambda _{19}+\frac{i a_2 \lambda _{19}}{10 \sqrt{6}}-\frac{1}{10} \sqrt{\frac{3}{2}} \lambda
_{21} \phi _1-\frac{\lambda _{21} \phi _2}{60 \sqrt{6}}-\frac{\lambda _{21} \phi _3}{6 \sqrt{30}},
\nonumber\\
m_{44}^{(3,2,\frac{1}{6})} &\equiv&
m_2-\frac{3 a_1 \lambda _6}{5 \sqrt{10}}+\frac{a_2
	\lambda _6}{10 \sqrt{15}}+\frac{\lambda _2 \phi _1}{5 \sqrt{15}}-\frac{\lambda _2 \phi _2}{60 \sqrt{15}}-\frac{\lambda _2 \phi _3}{30 \sqrt{3}},
\nonumber\\
m_{55}^{(3,2,\frac{1}{6})} &\equiv&
m_2+\frac{3 a_1 \lambda _6}{5 \sqrt{10}}-\frac{a_2 \lambda _6}{10 \sqrt{15}}+\frac{\lambda _2 \phi _1}{10 \sqrt{15}}-\frac{\lambda _2 \phi _2}{20\sqrt{15}},
\nonumber\\
m_{66}^{(3,2,\frac{1}{6})} &\equiv&
m_1+\frac{4}{5} \sqrt{\frac{2}{5}} a_1 \lambda _7-\frac{4 a_2 \lambda _7}{15\sqrt{15}}+\frac{E \lambda _{10}}{12 \sqrt{15}}+\frac{1}{2} \sqrt{\frac{3}{5}} \lambda _1 \phi _1-\frac{\lambda _1 \phi _2}{18 \sqrt{15}}-\frac{5 \lambda _1 \phi _3}{18 \sqrt{3}},
\nonumber\\
m_{67}^{(3,2,\frac{1}{6})} &\equiv&
-\frac{2}{3} \sqrt{\frac{2}{15}} a_2 \lambda _7+\frac{1}{3} \sqrt{\frac{5}{6}} E \lambda _{10}+\frac{1}{9}
\sqrt{\frac{5}{6}} \lambda _1 \phi _2-\frac{1}{9} \sqrt{\frac{2}{3}} \lambda _1 \phi _3,
\nonumber \\
m_{77}^{(3,2,\frac{1}{6})} &\equiv&
m_1-\frac{2}{5} \sqrt{\frac{2}{5}}
a_1 \lambda _7+\frac{22 a_2 \lambda _7}{15 \sqrt{15}}+\frac{11 E \lambda _{10}}{12 \sqrt{15}}-\frac{1}{18} \sqrt{\frac{5}{3}} \lambda _1
\phi _2+\frac{\lambda _1 \phi _3}{9 \sqrt{3}}.
\end{eqnarray}\\[0.2cm]
%
%
\begin{minipage}{16cm}
	$\left[{\bf{(1,2}}, \frac{1}{2}) +c.c. \right]$\\[.2cm]
	{\bf c:}
	$\widehat{H}^{(1,2,\frac{1}{2})}_{(5,2)}$,
	$\widehat{D}^{(1,2,\frac{1}{2})}_{(5,2)}$,
	$\widehat{D}^{(1,2,\frac{1}{2})}_{(45,2)}$,
	$\widehat{\Delta}^{(1,2,\frac{1}{2})}_{(45, 2)}$,
	$\widehat{\overline{\Delta}}^{(1,2,\frac{1}{2})}_{(5,2)}$,
	$\widehat{\Phi}^{(1,2,\frac{1}{2})}_{(5,-8)}$\\[.15cm]
	{\bf r:}
	$\widehat{H}^{(1,2,-\frac{1}{2})}_{(\overline{5},-2)}$,
	$\widehat{D}^{(1,2,-\frac{1}{2})}_{(\overline{5},-2)}$,
	$\widehat{D}^{(1,2,-\frac{1}{2})}_{(\overline{45},-2)}$,
	$\widehat{\overline{\Delta}}^{(1,2,-\frac{1}{2})}_{(\overline{45},-2)}$,
	$\widehat{\Delta}^{(1,2,-\frac{1}{2})}_{(\overline{5}, -2)}$,
	$\widehat{\Phi}^{(1,2,-\frac{1}{2})}_{(\overline{5},8)}$\\[.2cm]
	\begin{equation}
	\left(
	\begin{array}{cccccc}
	\begin{array}{c}
	m_3 + \sqrt{\frac{3}{5}} \lambda_{13} E \\
	m_{21}^{(1,2,\frac{1}{2})} \\
	m_{31}^{(1,2,\frac{1}{2})} \\
	- \frac{\lambda_{4} \phi_2}{2 \sqrt{3}}
	+ \frac{\lambda_{4} \phi_3}{\sqrt{15}} \\
	m_{15}^{(1,2,\frac{1}{2})} \\
	- \frac{\lambda_3 V_R}{\sqrt{5}}
	\end{array}
	\begin{array}{c}
	m_{12}^{(1,2,\frac{1}{2})} \\
	m_{22}^{(1,2,\frac{1}{2})} \\
	m_{23}^{(1,2,\frac{1}{2})} \\
	m_{42}^{(1,2,\frac{1}{2})} \\
	m_{52}^{(1,2,\frac{1}{2})} \\
	- \frac{\lambda_{20} V_R}{\sqrt{30}}
	\end{array}
	\begin{array}{c}
	m_{13}^{(1,2,\frac{1}{2})} \\
	m_{23}^{(1,2,\frac{1}{2})} \\
	m_{33}^{(1,2,\frac{1}{2})} \\
	m_{43}^{(1,2,\frac{1}{2})} \\
	m_{53}^{(1,2,\frac{1}{2})} \\
	0
	\end{array}
	\begin{array}{c}
	- \frac{\lambda_{3} \phi_2}{2 \sqrt{3}}
	+ \frac{\lambda_{3} \phi_3}{\sqrt{15}} \\
	m_{24}^{(1,2,\frac{1}{2})} \\
	m_{34}^{(1,2,\frac{1}{2})} \\
	m_{44}^{(1,2,\frac{1}{2})} \\
	\frac{\lambda_{11} E}{\sqrt{15}} \\
	0
	\end{array}
	\begin{array}{c}
	m_{15}^{(1,2,\frac{1}{2})} \\
	m_{25}^{(1,2,\frac{1}{2})} \\
	m_{35}^{(1,2,\frac{1}{2})} \\
	\frac{\lambda_{12} E}{\sqrt{15}} \\
	m_{55}^{(1,2,\frac{1}{2})} \\
	\frac{\lambda_2 V_R}{10}
	\end{array}
	\begin{array}{c}
	-\frac{\lambda_4 \overline{V_R}}{\sqrt{5}} \\
	-\frac{\lambda_{21} \overline{V_R}}{\sqrt{30}} \\
	0 \\
	0 \\
	\frac{\lambda_2 \overline{V_R}}{10} \\
	m_{66}^{(1,2,\frac{1}{2})}
	\end{array}
	\end{array}
	\right),	
	\end{equation}
\end{minipage}\\[.2cm]
where
\begin{align}
m_{12}^{(1,2,\frac{1}{2})} & \equiv
-\frac{2 i a_1 \lambda _{16}}{\sqrt{15}}+\frac{i a_2 \lambda _{16}}{2 \sqrt{10}}
-\frac{\lambda_{17} \phi _1}{\sqrt{10}}-\frac{3 \lambda _{17} \phi _2}{4 \sqrt{10}}, \notag \\
m_{13}^{(1,2,\frac{1}{2})} & \equiv
-\frac{1}{2} i \sqrt{\frac{5}{6}} a_2 \lambda _{16}-\frac{1}{4} \sqrt{\frac{5}{6}}
\lambda _{17} \phi _2-\frac{\lambda _{17} \phi _3}{\sqrt{6}}, \notag \\
m_{15}^{(1,2,\frac{1}{2})} & \equiv
-\frac{1}{5}
\sqrt{3} \lambda _4 \phi _1+\frac{1}{10} \sqrt{3} \lambda _4 \phi _2, \notag \\
m_{21}^{(1,2,\frac{1}{2})} & \equiv
\frac{2 i a_1 \lambda _{16}}{\sqrt{15}}-\frac{i a_2 \lambda _{16}}{2 \sqrt{10}}-\frac{\lambda _{17} \phi _1}{\sqrt{10}}-\frac{3 \lambda _{17} \phi_2}{4 \sqrt{10}}, \notag \\
m_{22}^{(1,2,\frac{1}{2})} & \equiv
m_6+\frac{\lambda_{14} E}{2 \sqrt{15}}+\frac{\lambda _{15} \phi _1}{\sqrt{15}}
- \frac{\lambda _{15} \phi _2}{2 \sqrt{15}}, \notag \\
m_{23}^{(1,2,\frac{1}{2})} & \equiv
\frac{1}{6} \sqrt{5} \lambda _{14} E +\frac{1}{18} \sqrt{5} \lambda _{15} \phi _2
- \frac{\lambda _{15} \phi _3}{9}, \notag \\
m_{24}^{(1,2,\frac{1}{2})} & \equiv
- \frac{i a_2 \lambda_{18}}{4 \sqrt{2}}+\frac{\lambda _{20} \phi _2}{24 \sqrt{2}}
+ \frac{\lambda _{20} \phi _3}{6 \sqrt{10}}, \notag \\
m_{25}^{(1,2,\frac{1}{2})} & \equiv
-\frac{1}{5} i \sqrt{3} a_1 \lambda _{19}+\frac{3 i a_2 \lambda _{19}}{20 \sqrt{2}}
+\frac{\lambda _{21} \phi _1}{10 \sqrt{2}} +\frac{3 \lambda _{21} \phi _2}{40 \sqrt{2}}, \notag \\
m_{31}^{(1,2,\frac{1}{2})} & \equiv
\frac{1}{2} i \sqrt{\frac{5}{6}} a_2 \lambda _{16}-\frac{1}{4} \sqrt{\frac{5}{6}} \lambda _{17} \phi _2-\frac{\lambda _{17} \phi _3}{\sqrt{6}}, \notag \\
m_{33}^{(1,2,\frac{1}{2})} & \equiv
m_6+\frac{13 \lambda_{14} E}{6 \sqrt{15}}-\frac{\lambda _{15} \phi _1}{3 \sqrt{15}}
-\frac{7 \lambda _{15} \phi _2}{18 \sqrt{15}}+\frac{2 \lambda _{15} \phi _3}{9 \sqrt{3}}, \notag \\
m_{34}^{(1,2,\frac{1}{2})} & \equiv
-\frac{1}{5} i a_1 \lambda _{18}+\frac{13 i a_2 \lambda _{18}}{20 \sqrt{6}}
+\frac{\lambda _{20} \phi _1}{10 \sqrt{6}}-\frac{\lambda _{20}
	\phi _2}{120 \sqrt{6}}-\frac{\lambda _{20} \phi _3}{3 \sqrt{30}}, \notag \\
m_{35}^{(1,2,\frac{1}{2})} & \equiv
\frac{i a_2 \lambda _{19}}{4 \sqrt{6}}-\frac{\lambda _{21} \phi _2}{24 \sqrt{6}}
-\frac{\lambda_{21} \phi _3}{6 \sqrt{30}}, \notag \\
m_{42}^{(1,2,\frac{1}{2})} & \equiv
\frac{i a_2 \lambda _{19}}{4 \sqrt{2}}+\frac{\lambda _{21} \phi _2}{24
	\sqrt{2}}+\frac{\lambda _{21} \phi _3}{6 \sqrt{10}}, \notag \\
m_{43}^{(1,2,\frac{1}{2})} & \equiv
\frac{1}{5} i a_1 \lambda _{19}-\frac{13 i a_2 \lambda _{19}}{20 \sqrt{6}}+\frac{\lambda _{21}
	\phi _1}{10 \sqrt{6}}-\frac{\lambda _{21} \phi _2}{120 \sqrt{6}}-\frac{\lambda _{21} \phi _3}{3 \sqrt{30}}, \notag \\
m_{44}^{(1,2,\frac{1}{2})} & \equiv
m_2+\frac{a_1 \lambda _6}{5 \sqrt{10}}+\frac{1}{10} \sqrt{\frac{3}{5}} a_2 \lambda _6
-\frac{\lambda _2 \phi _2}{6 \sqrt{15}}+\frac{\lambda _2 \phi _3}{15 \sqrt{3}}, \notag \\
m_{52}^{(1,2,\frac{1}{2})} & \equiv
\frac{1}{5} i \sqrt{3} a_1 \lambda _{18}-\frac{3 i a_2 \lambda_{18}}{20 \sqrt{2}}+
\frac{\lambda _{20} \phi _1}{10 \sqrt{2}}+\frac{3 \lambda _{20} \phi _2}{40 \sqrt{2}}, \notag \\
m_{53}^{(1,2,\frac{1}{2})} & \equiv
-\frac{i a_2 \lambda _{18}}{4 \sqrt{6}}-\frac{\lambda_{20} \phi _2}{24 \sqrt{6}}
-\frac{\lambda _{20} \phi _3}{6 \sqrt{30}}, \notag \\
m_{55}^{(1,2,\frac{1}{2})} & \equiv
m_2-\frac{a_1 \lambda _6}{5 \sqrt{10}}-\frac{1}{10} \sqrt{\frac{3}{5}} a_2 \lambda _6
+\frac{\lambda _2 \phi _1}{5 \sqrt{15}}-\frac{\lambda _2 \phi _2}{10 \sqrt{15}}, \notag \\
m_{66}^{(1,2,\frac{1}{2})} & \equiv
m_1+\frac{2}{5} \sqrt{\frac{2}{5}} a_1 \lambda_7+\frac{2}{5} \sqrt{\frac{3}{5}} a_2 \lambda _7
-\frac{1}{4} \sqrt{\frac{3}{5}} \lambda _{10} E +\sqrt{\frac{3}{5}} \lambda _1 \phi _1-\frac{1}{2}
\sqrt{\frac{3}{5}} \lambda _1 \phi _2 .
\end{align}
\begin{minipage}{16cm}
	$\left[{\bf{(3,1}}, -\frac{1}{3}) +c.c. \right]$\\[.2cm]
	{\bf c:}
	$\widehat{H}^{(3,1,-\frac{1}{3})}_{(5,2)}$,
	$\widehat{D}^{(3,1,-\frac{1}{3})}_{(5,2)}$,
	$\widehat{D}^{(3,1,-\frac{1}{3})}_{(45,2)}$,
	$\widehat{\Delta}^{(3,1,-\frac{1}{3})}_{(45, 2)}$,
	$\widehat{\overline{\Delta}}^{(3,1,-\frac{1}{3})}_{(5,2)}$,
	$\widehat{\overline{\Delta}}^{(3,1,-\frac{1}{3})}_{(50,2)}$,
	$\widehat{\Phi}^{(3,1,-\frac{1}{3})}_{(5,-8)}$\\[.15cm]
	{\bf r:}
	$\widehat{H}^{(\overline{3},1,\frac{1}{3})}_{(\overline{5},-2)}$,
	$\widehat{D}^{(\overline{3},1,\frac{1}{3})}_{(\overline{5},-2)}$,
	$\widehat{D}^{(\overline{3},1,\frac{1}{3})}_{(\overline{45},-2)}$,
	$\widehat{\overline{\Delta}}^{(\overline{3},1,\frac{1}{3})}_{(\overline{45},-2)}$,
	$\widehat{\Delta}^{(\overline{3},1,\frac{1}{3})}_{(\overline{5}, -2)}$,
	$\widehat{\Delta}^{(\overline{3},1,\frac{1}{3})}_{(\overline{50},-2)}$,
	$\widehat{\Phi}^{(\overline{3},1,\frac{1}{3})}_{(\overline{5},8)}$\\[.2cm]
	\begin{equation}
	\left(
	\begin{array}{ccccccc}
	\begin{array}{c}
	m_3 - \frac{2 \lambda_{13} E}{\sqrt{15}} \\
	m_{21}^{(3,1,-\frac{1}{3})} \\
	m_{31}^{(3,1,-\frac{1}{3})} \\
	m_{14}^{(3,1,-\frac{1}{3})} \\
	m_{15}^{(3,1,-\frac{1}{3})} \\
	-\sqrt{\frac{2}{15}} \lambda_3 \phi_3 \\
	\frac{\lambda_3 V_R}{\sqrt{5}}
	\end{array}
	\begin{array}{c}
	m_{12}^{(3,1,-\frac{1}{3})} \\
	m_{22}^{(3,1,-\frac{1}{3})} \\
	m_{23}^{(3,1,-\frac{1}{3})} \\
	m_{42}^{(3,1,-\frac{1}{3})} \\
	m_{52}^{(3,1,-\frac{1}{3})} \\
	0 \\
	\frac{\lambda_{20} V_R}{\sqrt{30}}
	\end{array}
	\begin{array}{c}
	m_{13}^{(3,1,-\frac{1}{3})} \\
	m_{23}^{(3,1,-\frac{1}{3})} \\
	m_{33}^{(3,1,-\frac{1}{3})} \\
	m_{43}^{(3,1,-\frac{1}{3})} \\
	m_{53}^{(3,1,-\frac{1}{3})} \\
	m_{63}^{(3,1,-\frac{1}{3})} \\
	0
	\end{array}
	\begin{array}{c}
	m_{14}^{(3,1,-\frac{1}{3})} \\
	m_{24}^{(3,1,-\frac{1}{3})} \\
	m_{34}^{(3,1,-\frac{1}{3})} \\
	m_{44}^{(3,1,-\frac{1}{3})} \\
	\frac{2 \lambda_{11} E}{3 \sqrt{5}} \\
	\frac{2}{3} \sqrt{\frac{2}{5}} \lambda_{11} E \\
	0
	\end{array}
	\begin{array}{c}
	m_{15}^{(3,1,-\frac{1}{3})} \\
	m_{25}^{(3,1,-\frac{1}{3})} \\
	m_{35}^{(3,1,-\frac{1}{3})} \\
	\frac{2 \lambda_{12} E}{3 \sqrt{5}} \\
	m_{55}^{(3,1,-\frac{1}{3})} \\
	- \frac{\lambda_2 \phi_3}{15 \sqrt{6}} \\
	-\frac{\lambda_{2} V_R}{10}
	\end{array}
	\begin{array}{c}
	-\sqrt{\frac{2}{15}} \lambda_4 \phi_3 \\
	0 \\
	m_{36}^{(3,1,-\frac{1}{3})} \\
	\frac{2}{3} \sqrt{\frac{2}{5}} \lambda_{12} E\\
	- \frac{\lambda_2 \phi_3}{15 \sqrt{6}} \\
	m_{66}^{(3,1,-\frac{1}{3})} \\
	0
	\end{array}
	\begin{array}{c}
	\frac{\lambda_4 \overline{V_R}}{\sqrt{5}} \\
	\frac{\lambda_{21} \overline{V_R}}{\sqrt{30}} \\
	0 \\
	0 \\
	-\frac{\lambda_{2} \overline{V_R}}{10} \\
	0 \\
	m_{77}^{(3,1,-\frac{1}{3})}
	\end{array}
	\end{array}
	\right),
	\end{equation}
\end{minipage}\\[.2cm]
where
\begin{align}
m_{12}^{(3,1,-\frac{1}{3})} & \equiv
-\frac{2 i a_1 \lambda _{16}}{\sqrt{15}}-\frac{i a_2 \lambda _{16}}{3 \sqrt{10}}-\frac{\lambda
	_{17} \phi _1}{\sqrt{10}}+\frac{\lambda _{17} \phi _2}{2 \sqrt{10}}, \notag \\
m_{13}^{(3,1,-\frac{1}{3})} & \equiv
-\frac{1}{3} i \sqrt{\frac{5}{2}} a_2 \lambda _{16}-\frac{1}{6} \sqrt{\frac{5}{2}}
\lambda _{17} \phi _2+\frac{\lambda _{17} \phi _3}{3 \sqrt{2}}, \notag \\
m_{14}^{(3,1,-\frac{1}{3})} & \equiv
-\frac{1}{3} \lambda _3 \phi _2-\frac{\lambda _3 \phi _3}{3 \sqrt{5}}, \notag \\
m_{15}^{(3,1,-\frac{1}{3})} & \equiv
-\frac{1}{5}
\sqrt{3} \lambda _4 \phi _1-\frac{\lambda _4 \phi _2}{5 \sqrt{3}}, \notag \\
m_{21}^{(3,1,-\frac{1}{3})} & \equiv
\frac{2 i a_1 \lambda _{16}}{\sqrt{15}}+\frac{i a_2 \lambda _{16}}{3 \sqrt{10}}-\frac{\lambda _{17} \phi _1}{\sqrt{10}}+\frac{\lambda _{17} \phi
	_2}{2 \sqrt{10}}, \notag \\
m_{22}^{(3,1,-\frac{1}{3})} & \equiv
m_6-\frac{\lambda _{14} E}{3 \sqrt{15}}+\frac{\lambda _{15} \phi _1}{\sqrt{15}}+\frac{\lambda _{15} \phi _2}{3 \sqrt{15}}, \notag \\
m_{23}^{(3,1,-\frac{1}{3})} & \equiv
\frac{1}{3} \sqrt{\frac{5}{3}} \lambda _{14} E +\frac{1}{9} \sqrt{\frac{5}{3}} \lambda _{15} \phi _2+\frac{\lambda _{15} \phi _3}{9 \sqrt{3}}, \notag \\
m_{24}^{(3,1,-\frac{1}{3})} & \equiv
-\frac{i a_2 \lambda _{18}}{2 \sqrt{6}}+\frac{\lambda _{20} \phi _2}{12 \sqrt{6}}-\frac{\lambda _{20} \phi _3}{6 \sqrt{30}}, \notag \\
m_{25}^{(3,1,-\frac{1}{3})} & \equiv
-\frac{1}{5} i \sqrt{3}
a_1 \lambda _{19}-\frac{i a_2 \lambda _{19}}{10 \sqrt{2}}+\frac{\lambda _{21} \phi _1}{10 \sqrt{2}}-\frac{\lambda _{21} \phi _2}{20 \sqrt{2}}, \notag \\
m_{31}^{(3,1,-\frac{1}{3})} & \equiv
\frac{1}{3} i \sqrt{\frac{5}{2}} a_2 \lambda _{16}-\frac{1}{6} \sqrt{\frac{5}{2}} \lambda _{17} \phi _2+\frac{\lambda _{17} \phi _3}{3 \sqrt{2}}, \notag \\
m_{33}^{(3,1,-\frac{1}{3})} & \equiv
m_6-\frac{\lambda _{14} E}{3 \sqrt{15}}-\frac{\lambda _{15} \phi _1}{3 \sqrt{15}}-\frac{\lambda _{15} \phi _2}{9 \sqrt{15}}+\frac{4 \lambda
	_{15} \phi _3}{9 \sqrt{3}},  \notag \\
m_{34}^{(3,1,-\frac{1}{3})} & \equiv
-\frac{1}{5} i a_1 \lambda _{18}-\frac{i a_2 \lambda _{18}}{10 \sqrt{6}}+\frac{\lambda _{20} \phi _1}{10 \sqrt{6}}-\frac{\lambda
	_{20} \phi _2}{20 \sqrt{6}}, \notag \\
m_{35}^{(3,1,-\frac{1}{3})} & \equiv
\frac{i a_2 \lambda _{19}}{6 \sqrt{2}}-\frac{\lambda _{21} \phi _2}{36 \sqrt{2}}+\frac{\lambda _{21} \phi _3}{18 \sqrt{10}}, \notag \\
m_{36}^{(3,1,-\frac{1}{3})} & \equiv
-\frac{1}{3} i a_2 \lambda _{19}+\frac{\lambda _{21} \phi _2}{18}-\frac{\lambda _{21} \phi _3}{9 \sqrt{5}}, \notag \\
m_{42}^{(3,1,-\frac{1}{3})} & \equiv
\frac{i a_2 \lambda _{19}}{2 \sqrt{6}}+\frac{\lambda _{21} \phi _2}{12 \sqrt{6}}-\frac{\lambda
	_{21} \phi _3}{6 \sqrt{30}}, \notag \\
m_{43}^{(3,1,-\frac{1}{3})} & \equiv
\frac{1}{5} i a_1 \lambda _{19}+\frac{i a_2 \lambda _{19}}{10 \sqrt{6}}+\frac{\lambda _{21} \phi _1}{10 \sqrt{6}}-\frac{\lambda
	_{21} \phi _2}{20 \sqrt{6}}, \notag \\
m_{44}^{(3,1,-\frac{1}{3})} & \equiv
m_2+\frac{a_1 \lambda _6}{5 \sqrt{10}}-\frac{a_2 \lambda _6}{5 \sqrt{15}}, \notag \\
m_{52}^{(3,1,-\frac{1}{3})} & \equiv
\frac{1}{5} i \sqrt{3} a_1 \lambda _{18}+\frac{i a_2 \lambda _{18}}{10
	\sqrt{2}}+\frac{\lambda _{20} \phi _1}{10 \sqrt{2}}-\frac{\lambda _{20} \phi _2}{20 \sqrt{2}}, \notag \\
m_{53}^{(3,1,-\frac{1}{3})} & \equiv
-\frac{i a_2 \lambda _{18}}{6 \sqrt{2}}-\frac{\lambda
	_{20} \phi _2}{36 \sqrt{2}}+\frac{\lambda _{20} \phi _3}{18 \sqrt{10}}, \notag \\
m_{55}^{(3,1,-\frac{1}{3})} & \equiv
m_2-\frac{a_1 \lambda _6}{5
	\sqrt{10}}+\frac{a_2 \lambda _6}{5 \sqrt{15}}+\frac{\lambda _2 \phi _1}{5 \sqrt{15}}+\frac{\lambda _2 \phi _2}{15 \sqrt{15}}, \notag \\
m_{63}^{(3,1,-\frac{1}{3})} & \equiv
\frac{1}{3} i a_2 \lambda _{18}+\frac{\lambda _{20} \phi _2}{18}-\frac{\lambda _{20} \phi _3}{9 \sqrt{5}}, \notag \\
m_{66}^{(3,1,-\frac{1}{3})} & \equiv
m_2-\frac{a_1 \lambda _6}{5 \sqrt{10}}+\frac{a_2
	\lambda _6}{5 \sqrt{15}}-\frac{\lambda _2 \phi _1}{10 \sqrt{15}}-\frac{\lambda _2 \phi _2}{30 \sqrt{15}}+\frac{\lambda _2 \phi _3}{15 \sqrt{3}}, \notag \\
m_{77}^{(3,1,-\frac{1}{3})} & \equiv
m_1+\frac{2}{5} \sqrt{\frac{2}{5}} a_1
\lambda _7-\frac{4 a_2 \lambda _7}{5 \sqrt{15}}+\frac{\lambda _{10} E}{2 \sqrt{15}}+\sqrt{\frac{3}{5}} \lambda _1 \phi _1+\frac{\lambda _1 \phi _2}{\sqrt{15}}.
\end{align}
%
%
\begin{minipage}{16cm}
	$\left[{\bf{(1,1}}, 2) +c.c. \right]$\\[.2cm]
	{\bf c:}
	$\widehat{\Delta}^{(1,1,2)}_{(\overline{50},-2)}$\\[.15cm]
	{\bf r:}
	$\widehat{\overline{\Delta}}^{(1,1,-2)}_{(50,2)}$\\[.2cm]
	\begin{equation}
	m_2
	- \frac{\lambda_6 a_1}{5\sqrt{10}}
	+ \frac{2}{5} \sqrt{\frac{3}{5}} \lambda_6 a_2
	- \frac{\lambda_2\phi_1}{10\sqrt{15}}
	- \frac{\lambda_2\phi_2}{5\sqrt{15}}
	+ \frac{\lambda_2\phi_3}{5\sqrt{3}}.
	\end{equation}
\end{minipage}\\[.2cm]
\begin{minipage}{16cm}
	$\left[{\bf{(1,2}}, \frac{3}{2}) +c.c. \right]$\\[.2cm]
	{\bf c:}
	$\widehat{\Phi}^{(1,2,\frac{3}{2})}_{(40,-4)}$\\[.15cm]
	{\bf r:}
	$\widehat{\Phi}^{(1,2,-\frac{3}{2})}_{(\overline{40},4)}$\\[.2cm]
	\begin{equation}
	m_1
	- \frac{1}{4} \sqrt{\frac{3}{5}} \lambda_{10} E
	- \frac{2}{5} \sqrt{\frac{2}{5}} \lambda_7 a_1
	- \frac{2}{5} \sqrt{\frac{3}{5}} \lambda_7 a_2
	- \frac{1}{2} \sqrt{\frac{5}{3}} \lambda_1 \phi_2
	+ \frac{\lambda_1 \phi_3}{\sqrt{3}}.
	\end{equation}
\end{minipage}\\[.2cm]
\begin{minipage}{16cm}
	${\bf{(1,3}}, 0)$\\[.2cm]
	{\bf c:}
	$\widehat{A}^{(1,3,0)}_{(24,0)}$,
	$\widehat{E}^{(1,3,0)}_{(24,0)}$,
	$\widehat{\Phi}^{(1,3,0)}_{(24,0)}$\\[.15cm]
	{\bf r:}
	$\widehat{A}^{(1,3,0)}_{(24,0)}$,
	$\widehat{E}^{(1,3,0)}_{(24,0)}$,
	$\widehat{\Phi}^{(1,3,0)}_{(24,0)}$\\[.2cm]
	\begin{equation}
	\left(
	\begin{array}{ccc}
	\begin{array}{c}
	m_{11}^{(1,3,0)} \\
	\sqrt{\frac{2}{5}} \lambda_9 a_1 + \sqrt{\frac{3}{5}} \lambda_9 a_2 \\
	m_{13}^{(1,3,0)}
	\end{array}
	\begin{array}{c}
	\sqrt{\frac{2}{5}} \lambda_9 a_1 + \sqrt{\frac{3}{5}} \lambda_9 a_2 \\
	m_5 + 3 \sqrt{\frac{3}{5}} \lambda_8 E \\
	m_{23}^{(1,3,0)}
	\end{array}
	\begin{array}{c}
	m_{13}^{(1,3,0)} \\
	m_{23}^{(1,3,0)} \\
	m_{33}^{(1,3,0)}
	\end{array}
	\end{array}
	\right),
	\end{equation}
\end{minipage}\\[.2cm]
where
\begin{align}
m_{11}^{(1,3,0)} & \equiv m_4 + \sqrt{\frac{3}{5}} \lambda_9 E - \frac{\lambda_5 \phi_1}{\sqrt{15}} - \frac{2 \lambda_5 \phi_2}{\sqrt{15}} - \frac{\lambda_5 \phi_3}{\sqrt{3}}, \notag \\
m_{13}^{(1,3,0)} & \equiv -\sqrt{\frac{2}{5}} \lambda_5 a_1 + \frac{2 \lambda_5 a_2}{\sqrt{15}} + \frac{2}{5} \sqrt{\frac{3}{5}} \lambda_7 \phi_1 - \frac{8 \lambda_7 \phi_2}{5 \sqrt{15}} + \frac{2 \lambda_7 \phi_3}{5 \sqrt{3}}, \notag \\
m_{23}^{(1,3,0)} & \equiv - \frac{1}{2} \sqrt{\frac{3}{5}} \lambda_{10} \phi_1 - \frac{\lambda_{10} \phi_2}{2 \sqrt{15}} + \frac{\lambda_{10} \phi_3}{2 \sqrt{3}}, \notag \\
m_{33}^{(1,3,0)} & \equiv m_1 + \frac{\lambda_{10} E}{2 \sqrt{15}} - \frac{4}{5} \sqrt{\frac{2}{5}} \lambda_7 a_1 + \frac{8 \lambda_7 a_2}{5 \sqrt{15}} +
\frac{\lambda_1 \phi_1}{2 \sqrt{15}} - \frac{7 \lambda_1 \phi_2}{3 \sqrt{15}} - \frac{\lambda_1 \phi_3}{6 \sqrt{3}}.
\end{align}
\begin{minipage}{16cm}
	$\left[{\bf{(1,3}}, 1) +c.c. \right]$\\[.2cm]
	{\bf c:}
	$\widehat{E}^{(1,3,1)}_{(15,4)}$,
	$\widehat{\overline{\Delta}}^{(1,3,1)}_{(15,-6)}$\\[.15cm]
	{\bf r:}
	$\widehat{E}^{(1,3,-1)}_{(\overline{15},-4)}$,
	$\widehat{\Delta}^{(1,3,-1)}_{(\overline{15},6)}$\\[.2cm]
	\begin{equation}
	\left(
	\begin{array}{cc}
	\begin{array}{c}
	m_5 + 3 \sqrt{\frac{3}{5}} \lambda_{8} E \\
	\frac{2 \lambda_{11} V_R}{5}
	\end{array}
	\begin{array}{c}
	\frac{2 \lambda_{12} \overline{V_R}}{5} \\
	m_2	+ \frac{3 \lambda_6 a_1}{5 \sqrt{10}} - \frac{1}{5} \sqrt{\frac{3}{5}} \lambda_6 a_2 + \frac{\lambda_2 \phi_1}{10 \sqrt{15}} - \frac{1}{10} \sqrt{\frac{3}{5}} \lambda_2 \phi_2
	\end{array}
	\end{array}
	\right).
	\end{equation}
\end{minipage}\\[.2cm]
\begin{minipage}{16cm}
	$\left[{\bf{(3,1}}, -\frac{4}{3}) +c.c. \right]$\\[.2cm]
	{\bf c:}
	$\widehat{D}^{(3,1,-\frac{4}{3})}_{(\overline{45},-2)}$,
	$\widehat{\overline{\Delta}}^{(3,1,-\frac{4}{3})}_{(\overline{45},-2)}$\\[.15cm]
	{\bf r:}
	$\widehat{\overline{D}}^{(\overline{3},1,\frac{4}{3})}_{(45,2)}$,
	$\widehat{\Delta}^{(\overline{3},1,\frac{4}{3})}_{(45,2)}$\\[.2cm]
	\begin{equation}
	\left(
	\begin{array}{cc}
	\begin{array}{c}
	m_{11}^{(3,1,-\frac{4}{3})} \\
	m_{21}^{(3,1,-\frac{4}{3})}
	\end{array}
	\begin{array}{c}
	m_{12}^{(3,1,-\frac{4}{3})} \\
	m_{22}^{(3,1,-\frac{4}{3})}
	\end{array}
	\end{array}
	\right),
	\end{equation}
\end{minipage}\\[.2cm]
where
\begin{align}
m_{11}^{(3,1,-\frac{4}{3})} & \equiv m_6 + \frac{4 \lambda_{14} E}{3 \sqrt{15}} - \frac{\lambda_{15} \phi_1}{3 \sqrt{15}} + \frac{4 \lambda_{15} \phi_2}{9 \sqrt{15}} + \frac{5 \lambda_{15} \phi_3}{9 \sqrt{3}}, \notag \\
m_{21}^{(3,1,-\frac{4}{3})} & \equiv \frac{i \lambda_{18} a_1}{5} - \frac{i}{5} \sqrt{\frac{2}{3}} \lambda_{18} a_2 - \frac{\lambda_{20} \phi_1}{10 \sqrt{6}}
- \frac{\lambda_{20} \phi_2}{5 \sqrt{6}} + \frac{\lambda_{20} \phi_3}{2 \sqrt{30}}, \notag \\
m_{12}^{(3,1,-\frac{4}{3})} & \equiv -\frac{i}{5} \lambda_{19} a_1 + \frac{i}{5} \sqrt{\frac{2}{3}} \lambda_{19} a_2 - \frac{\lambda_{21} \phi_1}{10 \sqrt{6}} - \frac{\lambda_{21} \phi_2}{5 \sqrt{6}} + \frac{\lambda_{21} \phi_3}{2 \sqrt{30}}, \notag \\
m_{22}^{(3,1,-\frac{4}{3})} & \equiv m_2 + \frac{\lambda_{6} a_1}{5 \sqrt{10}} + \frac{4 \lambda_{6} a_2}{5 \sqrt{15}} + \frac{\lambda_{2} \phi_3}{10 \sqrt{3}}.
\end{align}
\begin{minipage}{16cm}
	$\left[{\bf{(3,1}}, \frac{5}{3}) +c.c. \right]$\\[.2cm]
	{\bf c:}
	$\widehat{\Phi}^{(3,1,\frac{5}{3})}_{(75,0)}$\\[.15cm]
	{\bf r:}
	$\widehat{\Phi}^{(\overline{3},1,-\frac{5}{3})}_{(75,0)}$\\[.2cm]
	\begin{equation}
	m_1 + \frac{\lambda_{10} E}{2 \sqrt{15}}
	+ \frac{2}{5} \sqrt{\frac{2}{5}} \lambda_7 a_1
	- \frac{4 \lambda_{7} a_2}{5 \sqrt{15}}
	- \frac{\lambda_1 \phi_1}{\sqrt{15}}
	- \frac{\lambda_1 \phi_2}{3 \sqrt{15}}
	+ \frac{4 \lambda_1 \phi_3}{3 \sqrt{3}}.
	\end{equation}
\end{minipage}\\[.2cm]
\begin{minipage}{16cm}
	$\left[{\bf{(3,2}}, \frac{7}{6}) +c.c. \right]$\\[.2cm]
	{\bf c:}
	$\widehat{D}^{(3,2,\frac{7}{6})}_{(\overline{45}, -2)}$,
	$\widehat{\Delta}^{(3,2,\frac{7}{6})}_{(\overline{50}, -2)}$,
	$\widehat{\overline{\Delta}}^{(3,2,\frac{7}{6})}_{(\overline{45}, -2)}$\\[.15cm]
	{\bf r:}
	$\widehat{D}^{(\overline{3},2,-\frac{7}{6})}_{(45, 2)}$,
	$\widehat{\overline{\Delta}}^{(\overline{3},2,-\frac{7}{6})}_{(50, 2)}$,
	$\widehat{\Delta}^{(\overline{3},2,-\frac{7}{6})}_{(45, 2)}$\\[.2cm]
	\begin{equation}
	\left(
	\begin{array}{ccc}
	\begin{array}{c}
	m_{11}^{(3,2,\frac{7}{6})} \\
	m_{21}^{(3,2,\frac{7}{6})} \\
	m_{31}^{(3,2,\frac{7}{6})}
	\end{array}
	\begin{array}{c}
	m_{12}^{(3,2,\frac{7}{6})} \\
	m_{22}^{(3,2,\frac{7}{6})} \\
	\frac{\lambda_{11} E}{\sqrt{15}}
	\end{array}
	\begin{array}{c}
	m_{13}^{(3,2,\frac{7}{6})} \\
	\frac{\lambda_{12} E}{\sqrt{15}} \\
	m_{33}^{(3,2,\frac{7}{6})}
	\end{array}
	\end{array}
	\right),
	\end{equation}
\end{minipage}\\[.2cm]
where
\begin{align}
m_{11}^{(3,2,\frac{7}{6})} & \equiv m_6
- \frac{\lambda_{14} E}{3 \sqrt{15}}
- \frac{\lambda_{15}}{3 \sqrt{15}}
- \frac{2 \lambda_{15} \phi_2}{3 \sqrt{15}}
+ \frac{\lambda_{15} \phi_3}{3 \sqrt{3}}, \notag \\
m_{21}^{(3,2,\frac{7}{6})} & \equiv
\frac{i \lambda_{19} a_2}{2 \sqrt{6}}
- \frac{\lambda_{21} \phi_2}{12 \sqrt{6}}
+ \frac{1}{3} \sqrt{\frac{2}{15}} \lambda_{21} \phi_3, \notag \\
m_{31}^{(3,2,\frac{7}{6})} & \equiv
\frac{i \lambda_{18} a_1}{5}
+ \frac{i \lambda_{18} a_2}{10 \sqrt{6}}
- \frac{\lambda_{20} \phi_1}{10 \sqrt{6}}
+ \frac{13 \lambda_{20} \phi_2}{60 \sqrt{6}}
+ \frac{\lambda_{20} \phi_3}{6 \sqrt{30}} \notag \\
m_{12}^{(3,2,\frac{7}{6})} & \equiv
- \frac{i \lambda_{18} a_2}{2 \sqrt{6}}
- \frac{\lambda_{20} \phi_2}{12 \sqrt{6}}
+ \frac{1}{3} \sqrt{\frac{2}{15}} \lambda_{20} \phi_3, \notag \\
m_{22}^{(3,2,\frac{7}{6})} & \equiv m_2
- \frac{\lambda_{6} a_1}{5 \sqrt{10}}
+ \frac{7 \lambda_{6} a_2}{10 \sqrt{15}}
- \frac{\lambda_{2} \phi_1}{10 \sqrt{15}}
- \frac{7 \lambda_{2} \phi_2}{60 \sqrt{15}}
+ \frac{\lambda_{2} \phi_3}{15 \sqrt{3}}, \notag \\
m_{13}^{(3,2,\frac{7}{6})} & \equiv
- \frac{i \lambda_{19} a_1}{5}
- \frac{i \lambda_{19} a_2}{10 \sqrt{6}}
- \frac{\lambda_{21} \phi_1}{10 \sqrt{6}}
+ \frac{13 \lambda_{21} \phi_2}{60 \sqrt{6}}
+ \frac{\lambda_{21} \phi_3}{6 \sqrt{30}}, \notag \\
m_{33}^{(3,2,\frac{7}{6})} & \equiv m_2
+ \frac{\lambda_{6} a_1}{5 \sqrt{10}}
- \frac{7 \lambda_{6} a_2}{10 \sqrt{15}}
- \frac{\lambda_{2} \phi_2}{12 \sqrt{15}}
+ \frac{\lambda_{2} \phi_3}{30 \sqrt{3}}.	
\end{align}
\begin{minipage}{16cm}
	$\left[{\bf{(3,3}}, -\frac{1}{3}) +c.c. \right]$\\[.2cm]
	{\bf c:}
	$\widehat{D}^{(3,3,-\frac{1}{3})}_{(45,2)}$,
	$\widehat{\Delta}^{(3,3,-\frac{1}{3})}_{(45,2)}$\\[.15cm]
	{\bf r:}
	$\widehat{D}^{(\overline{3},3,\frac{1}{3})}_{(\overline{45},-2)}$
	$\widehat{\overline{\Delta}}^{(\overline{3},3,\frac{1}{3})}_
	{(\overline{45},-2)}$\\[.2cm]
	\begin{equation}
	\left(
	\begin{array}{cc}
	\begin{array}{c}
	m_{11}^{(3,3,-\frac{1}{3})} \\
	m_{21}^{(3,3,-\frac{1}{3})}
	\end{array}
	\begin{array}{c}
	m_{12}^{(3,3,-\frac{1}{3})} \\
	m_{22}^{(3,3,-\frac{1}{3})}	
	\end{array}
	\end{array}
	\right),
	\end{equation}
\end{minipage}\\[.2cm]
where
\begin{align}
m_{11}^{(3,3,-\frac{1}{3})} & \equiv m_6
+ \frac{4 \lambda_{14} E}{3 \sqrt{15}}
- \frac{\lambda_{15} \phi_1}{3 \sqrt{15}}
- \frac{2\lambda_{15} \phi_2}{3 \sqrt{15}}
- \frac{\lambda_{15} \phi_3}{3 \sqrt{3}}, \notag \\
m_{21}^{(3,3,-\frac{1}{3})} & \equiv
\frac{i \lambda_{19} a_1}{5}
- \frac{i}{5} \sqrt{\frac{2}{3}} \lambda_{19} a_2
+ \frac{\lambda_{21} \phi_1}{10 \sqrt{6}}
- \frac{1}{15} \sqrt{\frac{2}{3}} \lambda_{21} \phi_2
+ \frac{\lambda_{21} \phi_3}{6 \sqrt{30}}, \notag \\
m_{12}^{(3,3,-\frac{1}{3})} & \equiv
- \frac{i \lambda_{18} a_1}{5}
+ \frac{i}{5} \sqrt{\frac{2}{3}} \lambda_{18} a_2
+ \frac{\lambda_{20} \phi_1}{10 \sqrt{6}}
- \frac{1}{15} \sqrt{\frac{2}{3}} \lambda_{20} \phi_2
+ \frac{\lambda_{20} \phi_3}{6 \sqrt{30}}, \notag \\
m_{22}^{(3,3,-\frac{1}{3})} & \equiv m_2
+ \frac{\lambda_6 a_1}{5 \sqrt{10}}
- \frac{\lambda_6 a_2}{5 \sqrt{15}}
- \frac{\lambda_2 \phi_2}{6 \sqrt{15}}
- \frac{\lambda_2 \phi_3}{30 \sqrt{3}}.
\end{align}
\begin{minipage}{16cm}
	$\left[{\bf{(3,3}}, \frac{2}{3}) +c.c. \right]$\\[.2cm]
	{\bf c:}
	$\widehat{\Phi}^{(3,3,\frac{2}{3})}_{(40,-4)}$\\[.15cm]
	{\bf r:}
	$\widehat{\Phi}^{(\overline{3},3,-\frac{2}{3})}_{(\overline{40},4)}$\\[.2cm]
	\begin{equation}
	m_1
	+ \frac{\lambda_{10} E}{2 \sqrt{15}}
	- \frac{2}{5} \sqrt{\frac{2}{5}} \lambda_7 a_1
	+ \frac{4 \lambda_{7} a_2}{5 \sqrt{15}}
	- \frac{1}{3} \sqrt{\frac{5}{3}} \lambda_1 \phi_2
	- \frac{\lambda_1 \phi_3}{3 \sqrt{3}}.
	\end{equation}
\end{minipage}\\[.2cm]
\begin{minipage}{16cm}
	$\left[{\bf{(6,1}}, -\frac{2}{3}) +c.c. \right]$\\[.2cm]
	{\bf c:}
	$\widehat{E}^{(6,1,-\frac{2}{3})}_{(15,4)}$,
	$\widehat{\overline{\Delta}}^{(6,1,-\frac{2}{3})}_{(15,-6)}$\\[.15cm]
	{\bf r:}
	$\widehat{E}^{(\overline{6},1,\frac{2}{3})}_{(\overline{15},-4)}$,
	$\widehat{\Delta}^{(\overline{6},1,\frac{2}{3})}_{(\overline{15},6)}$\\[.2cm]
	\begin{equation}
	\left(
	\begin{array}{cc}
	\begin{array}{c}
	m_5 - 2 \sqrt{\frac{3}{5}} \lambda_{8} E \\
	\frac{2 \lambda_{11} V_R}{5}
	\end{array}
	\begin{array}{c}
	\frac{2 \lambda_{12} \overline{V_R}}{5} \\
	m_2
	+ \frac{3 \lambda_6 a_1}{5 \sqrt{10}}
	+ \frac{2 \lambda_6 a_2}{5 \sqrt{15}}
	+ \frac{\lambda_2 \phi_1}{10 \sqrt{15}}
	+ \frac{\lambda_2 \phi_2}{5 \sqrt{15}}	
	\end{array}
	\end{array}
	\right).
	\end{equation}
\end{minipage}\\[.2cm]
\begin{minipage}{16cm}
	$\left[{\bf{(6,1}}, \frac{1}{3}) +c.c. \right]$\\[.2cm]
	{\bf c:}
	$\widehat{D}^{(6,1,\frac{1}{3})}_{(\overline{45},-2)}$,
	$\widehat{\overline{\Delta}}^{(6,1,\frac{1}{3})}_{(\overline{45},-2)}$\\[.15cm]
	{\bf r:}
	$\widehat{D}^{(\overline{6},1,-\frac{1}{3})}_{(45,2)}$,
	$\widehat{\Delta}^{(\overline{6},1,-\frac{1}{3})}_{(45,2)}$\\[.2cm]
	\begin{equation}
	\left(
	\begin{array}{cc}
	\begin{array}{c}
	m_{11}^{(6,1,\frac{1}{3})} \\
	m_{21}^{(6,1,\frac{1}{3})}
	\end{array}
	\begin{array}{c}
	m_{12}^{(6,1,\frac{1}{3})} \\
	m_{22}^{(6,1,\frac{1}{3})}	
	\end{array}
	\end{array}
	\right),
	\end{equation}
\end{minipage}\\[.2cm]
where
\begin{align}
m_{11}^{(6,1,\frac{1}{3})} & \equiv
m_6 - \frac{2 \lambda_{14} E}{\sqrt{15}}
- \frac{\lambda_{15} \phi_1}{3 \sqrt{15}}
+ \frac{4 \lambda_{15} \phi_2}{9 \sqrt{15}}
- \frac{\lambda_{15} \phi_3}{9 \sqrt{3}}, \notag \\
m_{21}^{(6,1,\frac{1}{3})} & \equiv
\frac{i \lambda_{18} a_1}{5}
+ \frac{i}{5} \sqrt{\frac{3}{2}} \lambda_{18} a_2
- \frac{\lambda_{20} \phi_1}{10 \sqrt{6}}
- \frac{\lambda_{20} \phi_2}{30 \sqrt{6}}
+ \frac{\lambda_{20} \phi_3}{6 \sqrt{30}}, \notag \\
m_{12}^{(6,1,\frac{1}{3})} & \equiv
- \frac{i \lambda_{19} a_1}{5}
- \frac{i}{5} \sqrt{\frac{3}{2}} \lambda_{19} a_2
- \frac{\lambda_{21} \phi_1}{10 \sqrt{6}}
- \frac{\lambda_{21} \phi_2}{30 \sqrt{6}}
+ \frac{\lambda_{21} \phi_3}{6 \sqrt{30}}, \notag \\
m_{22}^{(6,1,\frac{1}{3})} & \equiv
m_2
+ \frac{\lambda_6 a_1}{5 \sqrt{10}}
- \frac{\lambda_6 a_2}{5 \sqrt{15}}
+ \frac{\lambda_2 \phi_2}{6 \sqrt{15}}
+ \frac{\lambda_2 \phi_3}{30 \sqrt{3}}.
\end{align}
\begin{minipage}{16cm}
	$\left[{\bf{(6,1}}, \frac{4}{3}) +c.c. \right]$\\[.2cm]
	{\bf c:}
	$\widehat{\overline{\Delta}}^{(6,1,\frac{4}{3})}_{(50,2)}$\\[.15cm]
	{\bf r:}
	$\widehat{\Delta}^{(\overline{6},1,-\frac{4}{3})}_{(\overline{50},-2)}$\\[.2cm]
	\begin{equation}
	m_2
	- \frac{\lambda_6 a_1}{5 \sqrt{10}}
	- \frac{4 \lambda_6 a_2}{5 \sqrt{15}}
	- \frac{\lambda_2 \phi_1}{10 \sqrt{15}}
	+ \frac{2 \lambda_2 \phi_2}{15 \sqrt{15}}
	+ \frac{\lambda_2 \phi_3}{15 \sqrt{3}}.
	\end{equation}
\end{minipage}\\[.2cm]
\begin{minipage}{16cm}
	$\left[{\bf{(6,2}}, -\frac{1}{6}) +c.c. \right]$\\[.2cm]
	{\bf c:}
	$\widehat{\Phi}^{(6,2,-\frac{1}{6})}_{(40,-4)}$\\[.15cm]
	{\bf r:}
	$\widehat{\Phi}^{(\overline{6},2,\frac{1}{6})}_{(\overline{40},4)}$\\[.2cm]
	\begin{equation}
	m_1
	- \frac{1}{4} \sqrt{\frac{3}{5}} \lambda_{10} E
	- \frac{2}{5} \sqrt{\frac{2}{5}} \lambda_7 a_1
	- \frac{2}{5} \sqrt{\frac{3}{5}} \lambda_7 a_2
	+ \frac{1}{6} \sqrt{\frac{5}{3}} \lambda_1 \phi_2
	- \frac{\lambda_1 \phi_3}{3 \sqrt{3}}.	
	\end{equation}
\end{minipage}\\[.2cm]
\begin{minipage}{16cm}
	$\left[{\bf{(6,2}}, \frac{5}{6}) +c.c. \right]$\\[.2cm]
	{\bf c:}
	$\widehat{\Phi}^{(6,2,\frac{5}{6})}_{(75,0)}$\\[.15cm]
	{\bf r:}
	$\widehat{\Phi}^{(\overline{6},2,-\frac{5}{6})}_{(75,0)}$\\[.2cm]
	\begin{equation}
	m_1
	- \frac{1}{4} \sqrt{\frac{3}{5}} \lambda_{10} E
	+ \frac{2}{5} \sqrt{\frac{2}{5}} \lambda_7 a_1
	+ \frac{2}{5} \sqrt{\frac{3}{5}} \lambda_7 a_2
	- \frac{\lambda_1 \phi_1}{\sqrt{15}}
	+ \frac{\lambda_1 \phi_2}{2 \sqrt{15}}.	
	\end{equation}
\end{minipage}\\[.2cm]
\begin{minipage}{16cm}
	$\left[{\bf{(6,3}}, \frac{1}{3}) +c.c. \right]$\\[.2cm]
	{\bf c:}
	$\widehat{\Delta}^{(6,3,\frac{1}{3})}_{(\overline{50},-2)}$\\[.15cm]
	{\bf r:}
	$\widehat{\overline{\Delta}}^{(\overline{6},3,-\frac{1}{3})}_{(50,2)}$\\[.2cm]
	\begin{equation}
	m_2
	- \frac{\lambda_6 a_1}{5 \sqrt{10}}
	+ \frac{\lambda_6 a_2}{5 \sqrt{15}}
	- \frac{\lambda_2 \phi_1}{10 \sqrt{15}}
	- \frac{\lambda_2 \phi_2}{30 \sqrt{15}}
	- \frac{\lambda_2 \phi_3}{15 \sqrt{3}}.
	\end{equation}
\end{minipage}\\[.2cm]
\begin{minipage}{16cm}
	${\bf{(8,1}}, 0)$\\[.2cm]
	{\bf c:}
	$\widehat{A}^{(8,1,0)}_{(24,0)}$,
	$\widehat{E}^{(8,1,0)}_{(24,0)}$,
	$\widehat{\Phi}^{(8,1,0)}_{(24,0)}$,
	$\widehat{\Phi}^{(8,1,0)}_{(75,0)}$\\[.15cm]
	{\bf r:}
	$\widehat{A}^{(8,1,0)}_{(24,0)}$,
	$\widehat{E}^{(8,1,0)}_{(24,0)}$,
	$\widehat{\Phi}^{(8,1,0)}_{(24,0)}$,
	$\widehat{\Phi}^{(8,1,0)}_{(75,0)}$\\[.2cm]
	\begin{equation}
	\left(
	\begin{array}{cccc}
	\begin{array}{c}
	m_{11}^{(8,1,0)} \\
	\sqrt{\frac{2}{5}} a_1 \lambda_9
	- \frac{2 a_2 \lambda_9}{\sqrt{15}} \\
	m_{13}^{(8,1,0)} \\
	m_{14}^{(8,1,0)}
	\end{array}
	\begin{array}{c}
	\sqrt{\frac{2}{5}} a_1 \lambda_9
	- \frac{2 a_2 \lambda_9}{\sqrt{15}} \\
	m_5-2 \sqrt{\frac{3}{5}} \lambda_8 E\\
	m_{23}^{(8,1,0)} \\
	\frac{1}{3} \sqrt{\frac{5}{6}} \lambda_{10} \phi_2
	- \frac{1}{3} \sqrt{\frac{2}{3}} \lambda_{10} \phi_3
	\end{array}
	\begin{array}{c}
	m_{13}^{(8,1,0)} \\
	m_{23}^{(8,1,0)} \\
	m_{33}^{(8,1,0)} \\
	m_{34}^{(8,1,0)}
	\end{array}
	\begin{array}{c}
	m_{14}^{(8,1,0)} \\
	\frac{1}{3} \sqrt{\frac{5}{6}} \lambda_{10} \phi_2
	- \frac{1}{3} \sqrt{\frac{2}{3}} \lambda_{10} \phi_3 \\
	m_{34}^{(8,1,0)} \\
	m_{44}^{(8,1,0)}
	\end{array}
	\end{array}
	\right),
	\end{equation}
\end{minipage}\\[.2cm]
where
\begin{align}
m_{11}^{(8,1,0)} & \equiv
m_4
- \frac{2 \lambda_9 E}{\sqrt{15}}
- \frac{\lambda_5 \phi_1}{\sqrt{15}}
+ \frac{4 \lambda_5 \phi_2}{3 \sqrt{15}}
- \frac{\lambda_5 \phi_3}{3 \sqrt{3}}, \notag \\
m_{13}^{(8,1,0)} & \equiv
- \sqrt{\frac{2}{5}} a_1 \lambda_5
- \frac{4 a_2 \lambda_5}{3 \sqrt{15}}
+ \frac{2}{5} \sqrt{\frac{3}{5}} \lambda_7 \phi_1
+ \frac{16 \lambda_7 \phi_2}{15 \sqrt{15}}
+ \frac{2 \lambda_7 \phi_3}{15 \sqrt{3}}, \notag \\
m_{14}^{(8,1,0)} & \equiv
\frac{1}{3} \sqrt{\frac{10}{3}} a_2 \lambda_5
+ \frac{2}{3} \sqrt{\frac{2}{15}} \lambda_7\phi_2
+ \frac{8}{15} \sqrt{\frac{2}{3}} \lambda_7 \phi_3, \notag \\
m_{23}^{(8,1,0)} & \equiv
-\frac{1}{2} \sqrt{\frac{3}{5}} \lambda_{10} \phi_1
+ \frac{\lambda_{10} \phi_2}{3 \sqrt{15}}
+ \frac{\lambda_{10} \phi_3}{6 \sqrt{3}}, \notag \\
m_{33}^{(8,1,0)} & \equiv
m_1
- \frac{4}{5} \sqrt{\frac{2}{5}} a_1 \lambda_7
- \frac{16 a_2 \lambda_7}{15 \sqrt{15}}
- \frac{\lambda_{10} E}{3 \sqrt{15}}
+ \frac{\lambda_1 \phi_1}{2 \sqrt{15}}
+ \frac{14 \lambda_1 \phi_2}{9 \sqrt{15}}
- \frac{\lambda_1 \phi_3}{18 \sqrt{3}}, \notag \\
m_{34}^{(8,1,0)} & \equiv
- \frac{2}{3} \sqrt{\frac{2}{15}} a_2 \lambda _7
- \frac{1}{3} \sqrt{\frac{5}{6}} \lambda_{10} E
- \frac{1}{9} \sqrt{\frac{5}{6}} \lambda_1 \phi_2
- \frac{2}{9} \sqrt{\frac{2}{3}} \lambda_1 \phi_3, \notag \\
m_{44}^{(8,1,0)} & \equiv
m_1
+ \frac{2}{5} \sqrt{\frac{2}{5}} \lambda_7 a_1
+ \frac{28 \lambda_7 a_2}{15 \sqrt{15}}
- \frac{7 \lambda_{10} E}{6 \sqrt{15}}
- \frac{\lambda_1 \phi_1}{\sqrt{15}}
+ \frac{7 \lambda_1 \phi_2}{9 \sqrt{15}}
+ \frac{2 \lambda_1 \phi_3}{9 \sqrt{3}}.
\end{align} \\[0.2cm]
\begin{minipage}{16cm}
	$\left[{\bf{(8,1}}, 1) +c.c. \right]$\\[.2cm]
	{\bf c:}
	$\widehat{\phi}^{(8,1,1)}_{(\overline{40},4)}$\\[.15cm]
	{\bf r:}
	$\widehat{\phi}^{(8,1,-1)}_{(40,-4)}$\\[.2cm]
	\begin{equation}
	m_1
	+ \frac{\lambda_{10} E}{2 \sqrt{15}}
	- \frac{2}{5} \sqrt{\frac{2}{5}} \lambda_7 a_1
	+ \frac{4 \lambda_7 a_2}{5 \sqrt{15}}
	+ \frac{1}{3} \sqrt{\frac{5}{3}} \lambda_1 \phi_2
	+ \frac{\lambda_1 \phi_3}{3 \sqrt{3}}.
	\end{equation}
\end{minipage}\\[.2cm]
\begin{minipage}{16cm}
	$\left[{\bf{(8,2}}, \frac{1}{2}) +c.c. \right]$\\[.2cm]
	{\bf c:}
	$\widehat{D}^{(8,2,\frac{1}{2})}_{(45,2)}$,
	$\widehat{\Delta}^{(8,2,\frac{1}{2})}_{(45,2)}$,
	$\widehat{\overline{\Delta}}^{(8,2,\frac{1}{2})}_{(50,2)}$\\[.15cm]
	{\bf r:}
	$\widehat{D}^{(8,2,-\frac{1}{2})}_{(\overline{45},-2)}$,
	$\widehat{\overline{\Delta}}^{(8,2,-\frac{1}{2})}_{(\overline{45},-2)}$,
	$\widehat{\Delta}^{(8,2,-\frac{1}{2})}_{(\overline{50},-2)}$\\[.2cm]
	\\[.2cm]
	\begin{equation}
	\left(
	\begin{array}{ccc}
	\begin{array}{c}
	m_{11}^{(8,2,\frac{1}{2})} \\
	m_{21}^{(8,2,\frac{1}{2})} \\
	m_{31}^{(8,2,\frac{1}{2})}
	\end{array}
	\begin{array}{c}
	m_{12}^{(8,2,\frac{1}{2})} \\
	m_{22}^{(8,2,\frac{1}{2})} \\
	\frac{\lambda_{11} E}{\sqrt{15}}
	\end{array}
	\begin{array}{c}
	m_{13}^{(8,2,\frac{1}{2})} \\
	\frac{\lambda_{12} E}{\sqrt{15}} \\
	m_{33}^{(8,2,\frac{1}{2})}
	\end{array}
	\end{array}
	\right),
	\end{equation}
\end{minipage}\\[.2cm]
where
\begin{align}
m_{11}^{(8,2,\frac{1}{2})} & \equiv
m_6-\frac{\lambda _{14} E}{3 \sqrt{15}}-\frac{\lambda _{15} \phi _1}{3 \sqrt{15}}+\frac{4 \lambda _{15} \phi _2}{9 \sqrt{15}}-\frac{\lambda
	_{15} \phi _3}{9 \sqrt{3}}, \notag \\
m_{12}^{(8,2,\frac{1}{2})} & \equiv
-\frac{1}{5} i a_1 \lambda _{18}-\frac{i a_2 \lambda _{18}}{10 \sqrt{6}}+\frac{\lambda _{20} \phi _1}{10 \sqrt{6}}+\frac{7
	\lambda _{20} \phi _2}{60 \sqrt{6}}+\frac{\lambda _{20} \phi _3}{6 \sqrt{30}}, \notag \\
m_{13}^{(8,2,\frac{1}{2})} & \equiv
-\frac{i a_2 \lambda _{19}}{2 \sqrt{6}}+\frac{\lambda _{21} \phi
	_2}{12 \sqrt{6}}+\frac{\lambda _{21} \phi _3}{3 \sqrt{30}}, \notag \\
m_{21}^{(8,2,\frac{1}{2})} & \equiv
\frac{1}{5} i a_1 \lambda _{19}+\frac{i a_2 \lambda _{19}}{10 \sqrt{6}}+\frac{\lambda _{21} \phi _1}{10 \sqrt{6}}+\frac{7 \lambda _{21} \phi _2}{60
	\sqrt{6}}+\frac{\lambda _{21} \phi _3}{6 \sqrt{30}}, \notag \\
m_{22}^{(8,2,\frac{1}{2})} & \equiv
m_2+\frac{a_1 \lambda _6}{5 \sqrt{10}}+\frac{1}{10} \sqrt{\frac{3}{5}} a_2 \lambda _6+\frac{\lambda
	_2 \phi _2}{12 \sqrt{15}}-\frac{\lambda _2 \phi _3}{30 \sqrt{3}}, \notag \\
m_{31}^{(8,2,\frac{1}{2})} & \equiv
\frac{i a_2 \lambda _{18}}{2 \sqrt{6}}+\frac{\lambda _{20} \phi _2}{12 \sqrt{6}}+\frac{\lambda _{20} \phi _3}{3 \sqrt{30}} \notag \\
m_{33}^{(8,2,\frac{1}{2})} & \equiv
m_2-\frac{a_1 \lambda _6}{5 \sqrt{10}}-\frac{1}{10} \sqrt{\frac{3}{5}} a_2 \lambda _6-\frac{\lambda _2 \phi _1}{10 \sqrt{15}}+\frac{\lambda
	_2 \phi _2}{20 \sqrt{15}}.
\end{align}
\begin{minipage}{16cm}
	${\bf{(8,3}}, 0)$\\[.2cm]
	{\bf c:}
	$\widehat{\Phi}^{(8,3,0)}_{(75,0)}$\\[.15cm]
	{\bf r:}
	$\widehat{\Phi}^{(8,3,0)}_{(75,0)}$\\[.2cm]
	\begin{equation}
	m_1
	+ \frac{\lambda_{10} E}{2 \sqrt{15}}
	+ \frac{2}{5} \sqrt{\frac{2}{5}} \lambda_7 a_1
	- \frac{4 \lambda_7 a_2}{5 \sqrt{15}}
	- \frac{\lambda_1 \phi_1}{\sqrt{15}}
	- \frac{\lambda_1 \phi_2}{3 \sqrt{15}}
	- \frac{2 \lambda_1 \phi_3}{3 \sqrt{3}}.
	\end{equation}
\end{minipage}

\section{Discussions}

The results presented in the above Sections can be compared with previous studies,
provided suitable transformations in notations are made. Firstly,
in a recent study\cite{nath2015}, $G_{51}$ is used in the MSSO10 with
$126 + \overline{126}+210$ breaking the SO(10).
Relations of the VEVs are
\begin{eqnarray}
\phi_1 &=& -\frac{1}{4}\sqrt{\frac{3}{5}} \mathbf{S}_{1_{210}}, \qquad
\phi_2 = \frac{1}{2} \sqrt{\frac{5}{3}} \mathbf{S}_{24_{210}}, \qquad
\phi_3 = \frac{\sqrt{3}}{2} \mathbf{S}_{75_{210}}, \nonumber \\
V_R &=& \sqrt{\frac{15}{2}} \mathbf{S}_{1_{126}}, \qquad
\overline{V_R} = \sqrt{\frac{15}{2}} \mathbf{S}_{1_{\overline{126}}}.
\end{eqnarray}
However, we find that the
coefficient before $\mathbf{S}_{75_{210}}^2 \mathbf{S}_{1_{210}}$ coupling should be $\frac{3}{160}$ instead
of $\frac{1}{40}$ in their (3).

We can also compare our results with \cite{fuku} which uses $G_{422}$ as the maximal subgroup.
When the following transformations are made,
	\begin{eqnarray} \left(
	\begin{array}{c}
	\hat{A}_{(1, 0)}^{(1, 1, 0)} \\
	\hat{A}_{(24, 0)}^{(1, 1, 0)}
	\end{array} \right)	= \left(
	\begin{array}{cc}
	\sqrt{\frac{2}{5}} & \sqrt{\frac{3}{5}}   \\
	\sqrt{\frac{3}{5}} & -\sqrt{\frac{2}{5}}
	\end{array} \right) \left(
	\begin{array}{c}
	\hat{A}_{(1, 1, 3)}^{(1, 1, 0)} \\
	\hat{A}_{(15, 1, 1)}^{(1, 1, 0)}
	\end{array} \right),
	\label{trans1}
	\end{eqnarray}

	\begin{eqnarray} \left(
	\begin{array}{c}
	\hat{\Phi}_{(1, 0)}^{(1, 1, 0)} \\
	\hat{\Phi}_{(24, 0)}^{(1, 1, 0)} \\
	\hat{\Phi}_{(75, 0)}^{(1, 1, 0)}
	\end{array} \right)	= \left(
	\begin{array}{ccc}
	\sqrt{\frac{1}{10}} & \sqrt{\frac{3}{10}} & \sqrt{\frac{6}{10}}  \\
	\sqrt{\frac{6}{15}} & -\sqrt{\frac{8}{15}} & \sqrt{\frac{1}{15}} \\
	\sqrt{\frac{3}{6}} & \sqrt{\frac{1}{6}} & -\sqrt{\frac{2}{6}}
	\end{array} \right) \left(
	\begin{array}{c}
	\hat{\Phi}_{(1, 1, 1)}^{(1, 1, 0)} \\
	\hat{\Phi}_{(15, 1, 1)}^{(1, 1, 0)} \\
	\hat{\Phi}_{(15, 1, 3)}^{(1, 1, 0)}
	\end{array} \right),
	\label{trans2}
	\end{eqnarray}

	\begin{eqnarray} \left(
	\begin{array}{c}
	\hat{\Phi}_{(\overline{10}, -4)}^{(3, 1, \frac{2}{3})} \\
	\hat{\Phi}_{(40, -4)}^{(3, 1, \frac{2}{3})}
	\end{array} \right)	= \left(
	\begin{array}{cc}
	\sqrt{\frac{1}{3}} & \sqrt{\frac{2}{3}}   \\
	\sqrt{\frac{2}{3}} & -\sqrt{\frac{1}{3}}
	\end{array} \right) \left(
	\begin{array}{c}
	\hat{\Phi}_{(15, 1, 1)}^{(3, 1, \frac{2}{3})} \\
	\hat{\Phi}_{(15, 1, 3)}^{(3, 1, \frac{2}{3})}
	\end{array} \right),
	\label{trans3}
	\end{eqnarray}

	\begin{eqnarray} \left(
	\begin{array}{c}
	\hat{\Phi}_{(24, 0)}^{(3, 2, -\frac{5}{6})} \\
	\hat{\Phi}_{(75, 0)}^{(3, 2, -\frac{5}{6})}
	\end{array} \right)	= \left(
	\begin{array}{cc}
	\sqrt{\frac{1}{3}} & \sqrt{\frac{2}{3}}   \\
	\sqrt{\frac{2}{3}} & -\sqrt{\frac{1}{3}}
	\end{array} \right) \left(
	\begin{array}{c}
	\hat{\Phi}_{(6, 2, 2)}^{(3, 2, -\frac{5}{6})} \\
	\hat{\Phi}_{(10, 2, 2)}^{(3, 2, -\frac{5}{6})}
	\end{array} \right),
	\label{trans4}
	\end{eqnarray}

	\begin{eqnarray} \left(
	\begin{array}{c}
	\hat{\Phi}_{(10, 4)}^{(3, 2, \frac{1}{6})} \\
	\hat{\Phi}_{(\overline{40}, 4)}^{(3, 2, \frac{1}{6})}
	\end{array} \right)	= \left(
	\begin{array}{cc}
	\sqrt{\frac{1}{3}} & \sqrt{\frac{2}{3}}   \\
	\sqrt{\frac{2}{3}} & -\sqrt{\frac{1}{3}}
	\end{array} \right) \left(
	\begin{array}{c}
	\hat{\Phi}_{(6, 2, 2)}^{(3, 2, \frac{1}{6})} \\
	\hat{\Phi}_{(10, 2, 2)}^{(3, 2, \frac{1}{6})}
	\end{array} \right),
	\label{trans5}
	\end{eqnarray}

	\begin{eqnarray} \left(
	\begin{array}{c}
	\hat{D}_{(5, 2)}^{(1, 2, \frac{1}{2})} \\
	\hat{D}_{(45, 2)}^{(1, 2, \frac{1}{2})}
	\end{array} \right)	= \left(
	\begin{array}{cc}
	\sqrt{\frac{1}{4}} & \sqrt{\frac{3}{4}}   \\
	\sqrt{\frac{3}{4}} & -\sqrt{\frac{1}{4}}
	\end{array} \right) \left(
	\begin{array}{c}
	\hat{D}_{(1, 2, 2)}^{(1, 2, \frac{1}{2})} \\
	\hat{D}_{(15, 2, 2)}^{(1, 2, \frac{1}{2})}
	\end{array} \right),
	\end{eqnarray}
	\label{trans6}

	\begin{eqnarray} \left(
	\begin{array}{c}
	\hat{D}_{(5, 2)}^{(3, 1, -\frac{1}{3})} \\
	\hat{D}_{(45, 2)}^{(3, 1, -\frac{1}{3})}
	\end{array} \right)	= \left(
	\begin{array}{cc}
	\sqrt{\frac{1}{2}} & \sqrt{\frac{1}{2}}   \\
	\sqrt{\frac{1}{2}} & -\sqrt{\frac{1}{2}}
	\end{array} \right) \left(
	\begin{array}{c}
	\hat{D}_{(6, 1, 3)}^{(3, 1, -\frac{1}{3})} \\
	\hat{D}_{(10, 1, 1)}^{(3, 1, -\frac{1}{3})}
	\end{array} \right),
	\end{eqnarray}
	\label{trans7}

	\begin{eqnarray} \left(
	\begin{array}{c}
	\hat{\overline{\Delta}}_{(5, 2)}^{(3, 1, -\frac{1}{3})} \\
	\hat{\overline{\Delta}}_{(50, 2)}^{(3, 1, -\frac{1}{3})}
	\end{array} \right)	= \left(
	\begin{array}{cc}
	\sqrt{\frac{1}{3}} & \sqrt{\frac{2}{3}}   \\
	\sqrt{\frac{2}{3}} & -\sqrt{\frac{1}{3}}
	\end{array} \right) \left(
	\begin{array}{c}
	\hat{\overline{\Delta}}_{(6, 1, 1)}^{(3, 1, -\frac{1}{3})} \\
	\hat{\overline{\Delta}}_{(10, 1, 3)}^{(3, 1, -\frac{1}{3})}
	\end{array} \right),
	\label{trans9}
	\end{eqnarray}

	\begin{eqnarray} \left(
	\begin{array}{c}
	\hat{\Phi}_{(24, 0)}^{(8, 1, 0)} \\
	\hat{\Phi}_{(75, 0)}^{(8, 1, 0)}
	\end{array} \right)	= \left(
	\begin{array}{cc}
	\sqrt{\frac{1}{3}} & \sqrt{\frac{2}{3}}   \\
	\sqrt{\frac{2}{3}} & -\sqrt{\frac{1}{3}}
	\end{array} \right) \left(
	\begin{array}{c}
	\hat{\Phi}_{(15, 1, 1)}^{(8, 1, 0)} \\
	\hat{\Phi}_{(15, 1, 3)}^{(8, 1, 0)}
	\end{array} \right),
	\label{trans10}
	\end{eqnarray}
\noindent our masses and mass matrices agree with those given in \cite{fuku},
excepts in \cite{fuku} the coefficients of 24 and 42 elements in the [(1, 1, 1) + c.c] mass matrix
should be $-\frac{1}{2\sqrt{10}}$ instead of $-\frac{1}{\sqrt{10}}$.

Furthermore, as is pointed out in \cite{czy},
the not all of dimensionless couplings in the superpotential (\ref{potential}) can be taken as order 1
in numerical calculations.
This can be seen in the the conventional SO(10) basis of $1'. 2', \cdots, 9', 0'$ where
the fields are defined as
\begin{eqnarray}
	\widehat{H} &\equiv& a' \qquad \widehat{A} \equiv \frac{1}{\sqrt{2}}(a'b') \qquad
	\widehat{E} \equiv \frac{1}{\sqrt{2}} \{a'b'\} \qquad
	\widehat{D} \equiv \frac{1}{\sqrt{3}}(a'b'c')  \nonumber \\
	\widehat{\Phi} &\equiv& \frac{1}{\sqrt{24}} (a'b'c'd') \qquad
	\widehat{\overline{\Delta}} \equiv \frac{1}{\sqrt{120}}(a'b'c'd'e'),
\end{eqnarray}
and in \cite{fuku} the SO(10) invariants are defined as, {\it e.g.},
\begin{eqnarray}
	\lambda_1 \Phi^3
	&\equiv& \lambda_1(\Phi_{a'b'c'd'} \widehat{\Phi}_{a'b'c'd'}) \cdot (\Phi_{a'b'e'f'} \widehat{\Phi}_{a'b'e'f'}) \cdot (\Phi_{c'd'e'f'} \widehat{\Phi}_{c'd'e'f'})  \nonumber \\
	&\equiv& \lambda_1\Phi_{a'b'c'd'} \Phi_{a'b'e'f'} \Phi_{c'd'e'f'} (\widehat{\Phi}_{a'b'c'd'} \cdot \widehat{\Phi}_{a'b'e'f'}  \cdot \widehat{\Phi}_{c'd'e'f'}) \nonumber  \\
	&=& \lambda_1 \Phi_{a'b'c'd'} \Phi_{a'b'e'f'} \Phi_{c'd'e'f'} \times  3! \times (\frac{1}{\sqrt{24}})^3 \times (a'b'c'd')\cdot(a'b'e'f') \cdot(c'd'e'f') \nonumber \\
	&=& \lambda_1 \Phi_{a'b'c'd'} \Phi_{a'b'e'f'} \Phi_{c'd'e'f'} \times  3! \times (\frac{1}{\sqrt{24}})^3 \times (2!)^3 \nonumber \\
	&=& \frac{1}{\sqrt{6}} \lambda_1 \Phi_{a'b'c'd'} \Phi_{a'b'e'f'} \Phi_{c'd'e'f'} \equiv \lambda'_1 \Phi_{a'b'c'd'} \Phi_{a'b'e'f'} \Phi_{c'd'e'f'},
\end{eqnarray}
where the $3!$ comes from the exchange symmetry of the three $\Phi$s,
and $2!^3$ comes from index constrictions.
So we need to redefine the couplings to absorb the extra factor $\frac{1}{\sqrt{6}}$.
We give the redefined couplings  in Table \ref{NewCouplings}.
When these new couplings are taken as order 1, and all the masses and  the VEVs
in the superpotential (\ref{potential})
are taken at the same scale, all terms in (\ref{VEVsCG}) are of the same order.

\begin{table}\caption{Normalized  couplings $\lambda'$s which are order one.}\label{NewCouplings}
	\begin{center}
		\begin{tabular}{|c||c|c|c|c|c|c|c|c|c|c|c|c|c|c|c|c|}
			\hline  \hline
			Old & $\lambda_1$ &$\lambda_2$ & $\lambda_3$ &  $\lambda_4$   & $\lambda_5$ &$\lambda_6$ & $\lambda_7$ &$\lambda_8$&$\lambda_9$& $\lambda_{10}$ & $\lambda_{11}$ 	   \\ \hline
			New & $\sqrt{6}\lambda'_1$ & $10\sqrt{6}\lambda'_2$ & $\sqrt{5}\lambda'_3$  &$\sqrt{5}\lambda'_4$& $\frac{\sqrt{6}}{2}\lambda'_5$   & $5\sqrt{2}\lambda'_6$&$\frac{5\sqrt{2}}{4}\lambda'_7$ & $\frac{\sqrt{2}}{3}\lambda'_8$ & $\sqrt{2}\lambda'_9$ & $2\sqrt{2}\lambda'_{10}$ & $\frac{5\sqrt{2}}{2}\lambda'_{11}$ \\
			\hline \hline
		\end{tabular}
	\end{center}
\end{table}
\begin{table}
	\begin{center}
		\begin{tabular}{|c||c|c|c|c|c|c|c|c|c|c|c|c}
			\hline  \hline
			Old &$\lambda_{12}$ & $\lambda_{13}$& $\lambda_{14}$& $\lambda_{15}$& $\lambda_{16}$ &$\lambda_{17}$ & $\lambda_{18}$ &  $\lambda_{19}$ & $\lambda_{20}$ & $\lambda_{21}$    \\ \hline
			New & $\frac{5\sqrt{2}}{2}\lambda'_{12}$ & $\sqrt{2}\lambda'_{13}$ & $\frac{3\sqrt{2}}{2}\lambda'_{14}$ & $\frac{3\sqrt{6}}{2}\lambda'_{15}$ &  $\sqrt{3}\lambda'_{16}$ & $2\lambda'_{17}$ & $\sqrt{10}\lambda'_{18}$ & $\sqrt{10}\lambda'_{19}$ & $2\sqrt{30}\lambda'_{20}$  & $2\sqrt{30}\lambda'_{21}$  \\
			\hline \hline
		\end{tabular}
	\end{center}
\end{table}

\section{Summary}

In summary, we have studied the renormalizable SUSY SO(10) model which contains
many important Higgs fields used previously.
We use $G_{51}$ as the maximal subgroup and our results are highly consistent with those given in $G_{422}$ basis.
The results of this work can be used for reference in model building.

We thank Z.-X. Ren for many discussions. ZYC also thanks Feng-Shu Jin for many helps.

\end{document}